\documentclass[lettersize,journal]{IEEEtran}
\usepackage{graphicx}
\usepackage{amsmath}
\usepackage{amssymb}
\usepackage{array}
\usepackage[caption=false,font=normalsize,labelfont=rm,textfont=rm]{subfig}
\usepackage{textcomp}
\usepackage{stfloats}
\usepackage{url}
\usepackage{verbatim}
\usepackage{algorithm}
\usepackage{tabularx}
\usepackage{bm}
\usepackage{tikz}
\usepackage[T1]{fontenc}
\usepackage{caption}
\usepackage[dvipsnames,table]{xcolor}
\usepackage{booktabs}
\usepackage{algpseudocode}
\usepackage{flushend}
\usepackage{makecell}
\usepackage{multirow}
\usepackage{balance}
\usepackage[table,xcdraw]{xcolor}
\def\BibTeX{{\rm B\kern-.05em{\sc i\kern-.025em b}\kern-.08em
		T\kern-.1667em\lower.7ex\hbox{E}\kern-.125emX}}
\usepackage{balance}
\usepackage{enumitem}
\makeatletter
\providecommand{\insert@pcolumn}{\insert@column}
\makeatother
\begin{document}
\title{Ada-TokenCom: Rate-Adaptive Token Communications via 
Large-Model-Driven Token Compression and Generation}
\author{Zijun Zhang, Li Qiao,~\IEEEmembership{Member,~IEEE}, Mahdi Boloursaz Mashhadi,~\IEEEmembership{Senior Member,~IEEE}, \\Zhen Gao,~\IEEEmembership{Senior Member,~IEEE}, Mehdi Bennis,~\IEEEmembership{Fellow,~IEEE}, and Kaibin Huang,~\IEEEmembership{Fellow,~IEEE}
\thanks{Z. Zhang and Z. Gao are with the School of
Information and Electronics, Beijing Institute of Technology (Zhuhai), Zhuhai 519088,
China (e-mail:\{meshinazusa, gaozhen16\}@bit.edu.cn); L. Qiao and K. Huang are with the Department of Electrical and Computer Engineering, The University of Hong Kong, Pokfulam Road, Hong Kong SAR (e-mail: \{qiaoli, huangkb\}@hku.hk); Mahdi Boloursaz Mashhadi is with the University of Surrey, Guildford, U.K. (email: m.boloursazmashhadi@surrey.ac.uk). Mehdi Bennis is with the Centre for Wireless Communications, University of Oulu, 90014 Oulu, Finland (e-mail: mehdi.bennis@oulu.fi). Corresponding authors: L. Qiao and Z. Gao.}
}

\maketitle
\begin{abstract}

Token Communications (TokenCom) has recently emerged as a new paradigm in which tokens serve as unified units for communication and computation, enabling efficient multimodal semantic and goal-oriented transmission. In this paper, we develop Ada-TokenCom, a rate-adaptive TokenCom framework based on large autoregressive models, which integrates next-token prediction with arithmetic coding to achieve ultra-low bitrate 
semantic communication at the token level. We propose a mixed 
reconstruction/generation scheme, where the transmitter encodes and transmits the highly informative tokens at the beginning of the token sequence leveraging a pre-trained autoregressive large model, while the receiver uses an identical model to predict the rest. Moreover, we design a Lyapunov-based algorithm to dynamically optimize both the source compression rate and the modulation and coding scheme, adapting to time-varying network conditions. Simulation results demonstrate that 
our proposed Ada-TokenCom framework outperforms both digital and 
deep joint source-channel coding-based semantic communication baselines. 

\end{abstract}
\vspace{-2mm}

\vspace{-5mm}
\section{Introduction}

The rapid progress of generative foundation models (GFMs) and multimodal large language models (MLLMs) is opening up new design opportunities for semantic communications. 
Modern foundation models increasingly process heterogeneous modalities through token sequences, where tokens may be discrete indices, latent visual codes, audio units, or continuous token embeddings depending on the modality and model architecture. 
Such tokenized representations provide a structured interface between communication and computation, and have recently motivated the development of token communication (TokenCom) frameworks~\cite{qiao2025todma,qiao2025tokencom}. 
In TokenCom, tokens serve as the basic units for semantic-level coding, transmission, inference, and generation, enabling communication systems to exploit the contextual and generative priors learned by pretrained GFMs and MLLMs.

\subsection{From Semantic Communications to Token Communications}

Semantic communication shifts wireless system design from bit-level recovery to semantic-level information exchange under bandwidth constraints and unreliable channels~\cite{SemCom1,SemCom2}. Existing methods typically rely on end-to-end learned joint source-channel coding and perform well for specific tasks and modalities~\cite{bourtsoulatze2019deep,wang2023wireless}. However, their learned representations are often coupled with particular modalities, tasks, datasets, or channel conditions, limiting generalization and complicating integration with practical digital wireless systems.

Recent advances in GFMs and MLLMs provide a different route based on tokenized representations. By converting heterogeneous source signals into structured token sequences, foundation models can process, predict, and generate multimodal content through a unified sequence-modeling interface~\cite{wangMultimodalLearningNexttoken2026,yin2024survey}. 
This observation has motivated the development of TokenCom, where tokens serve as the basic units for semantic-level coding, transmission, inference, and generation~\cite{qiao2025todma,qiao2025tokencom}. 
Among different token forms, discrete tokens are particularly attractive for digital semantic communication, since they can be naturally combined with entropy coding, channel coding, modulation, and link adaptation.

The token-based representation also changes how semantic information can be compressed and recovered. 
Instead of transmitting all source samples or continuous latent features, a transmitter may send only selected tokens or a compressed token sequence, while the receiver reconstructs or generates the remaining semantic content using pretrained generative priors. 
This possibility is closely related to the predictive capability of modern foundation models. 
Autoregressive (AR) models estimate the conditional distribution of each token given its preceding context, and such conditional probabilities can be used for entropy coding. 
When the predicted token probabilities are accurate, predictable tokens require shorter codewords, whereas uncertain tokens require more explicit information to be transmitted~\cite{deletang2024language,ren2025separate}. 
Therefore, AR token modeling provides a principled way to exploit context-induced predictability beyond conventional signal-level redundancy. Moreover, emerging architectures \cite{miwa2026onedpiece, tian2024visual}
have redefined AR modeling by replacing traditional raster-scan sequences with hierarchical, scale-aware generation. By leveraging scale-agnostic priors, these emerging architectures can robustly preserve essential global features, including spatial topology, environmental context, and color distributions, even when only a minimal set of critical tokens is transmitted. 

\subsection{Related Works}
\subsubsection{Generative Semantic Communication}

Generative semantic communication (GenSemCom) has recently attracted increasing attention by exploiting the generative AI priors for semantic reconstruction.
Early learned semantic communication systems are mainly built upon deep joint source-channel coding (DeepJSCC), where source signals are mapped to continuous channel symbols and reconstructed from noisy channel observations~\cite{bourtsoulatze2019deep,10589474}. 
DeepJSCC-based schemes can jointly optimize source representation and channel protection, while digital variants further quantize semantic features into bits to improve compatibility with practical communication modules~\cite{huang2025d2jscc}. 
Nevertheless, these methods are usually trained for specific modalities, tasks, or channel distributions.

Recent GenSemCom studies move beyond conventional DeepJSCC by incorporating diffusion models, GFMs, and MLLMs into the semantic reconstruction process. 
Diffusion-model-assisted methods have been developed to improve perceptual reconstruction quality and robustness under degraded channel conditions~\cite{10960324,zhang2026semantics,huang2025visual}. 
Foundation-model-based GenSemCom further exploits pretrained generative priors for low-rate image or video communication, where compact semantic descriptions, prompts, tokens, or partial representations are transmitted and the receiver uses generative models to recover the missing semantic content~\cite{10599525,10960413,11370276,11112664,10982132}. 
These studies demonstrate that generative priors can substantially reduce the amount of explicitly transmitted information.

Adaptive mechanisms have also been investigated in learned semantic communication systems, for example through variable-rate learned source-channel coding, latent masking, compression-ratio adaptation, or model-configuration selection~\cite{dai2022nonlinear,chen2025communication}. 
These methods provide important insights into adaptive semantic transmission, but their rate adaptation is mainly tied to continuous learned representations or preconfigured semantic models. 
This motivates us to conceive a rate-adaptive token communications (Ada-TokenCom) framework, and it can perform rate adaptation at the discrete token-sequence level, where the truncation length directly determines how many tokens are explicitly compressed and transmitted and how many tokens are generated at the receiver. 
This enables token compression, receiver-side prediction, and MCS selection to be jointly optimized under long-term wireless resource constraints.

\subsubsection{Recent Advances of TokenCom}

Existing TokenCom studies can be broadly categorized according to their design objectives: token-level reconstruction and protection, networked token transmission, and task-oriented edge inference. 
For source reconstruction, several works exploit token-level context to reduce transmission overhead or improve robustness against token loss. 
For example, loss-resilient image compression has been studied by recovering missing or corrupted latent tokens through masked visual token modeling~\cite{wang2025resicomp}. 
An information bottleneck perspective has also been introduced to learn compact token representations that preserve task- or semantics-relevant information~\cite{wei2026token}. 
Other studies improve token transmission reliability or efficiency through communication-aware token organization and protection. 
For instance, semantic packet aggregation groups correlated tokens to reduce packetization overhead and mitigate packet-loss impact~\cite{lee2025low}, while unequal error protection has been designed according to the non-uniform semantic importance of tokens~\cite{zhang2026tokencomuep}. 
Joint semantic-channel coding and modulation further maps token representations to channel symbols for reliable wireless transmission~\cite{ying2026joint}. 
For video and multimodal sources, textual intent and multimodal context have been used to guide multi-rate token transmission and adaptive source-channel protection~\cite{men2026video,zhang2026task}. 
These studies demonstrate that token-level context and semantic importance are useful for improving reconstruction robustness and reducing communication cost.

TokenCom has also been extended to multi-user and networked wireless scenarios. 
One line of work studies tokenizer agreement among multiple users, where transmitters and receivers need to select compatible tokenizers and codebooks under wireless resource constraints~\cite{zeinali2026wireless}. 
Another line exploits semantic context and inter-source semantic orthogonality to mitigate collisions among simultaneously transmitted token sequences~\cite{qiao2025todma, ToDMA}. 
These works show that tokens can serve not only as source representations, but also as network-level units for packetization, coordination, protection, and multiple access. 
However, their focus is mainly on tokenizer agreement, token-domain collision mitigation, or transmission reliability, rather than on rate-adaptive token compression and prediction over time-varying wireless channels.

Beyond reconstruction-oriented transmission, TokenCom has recently been studied as an interface for edge AI services. 
For example, vision-language token communication has been investigated to support multimodal and multitask inference at the receiver~\cite{jiang2026tokencom}. 
Task-relevant semantic token selection and wireless resource allocation have also been jointly considered for transformer-based edge inference under dynamic system conditions~\cite{devoto2026adaptive}. 
In addition, token merging has been used to reduce the number of transmitted tokens by combining redundant token representations while preserving edge inference performance~\cite{erak2026adaptive}. 
This line of work highlights the potential of TokenCom for connecting wireless transmission with downstream MLLM inference, rather than merely reconstructing source signals.

Overall, existing studies have established key mechanisms for token recovery, selection, protection, packetization, tokenizer agreement, multiple access, and task-oriented token transmission. 
Building on these foundations, our proposed Ada-TokenCom focuses on the adaptive control of digital token transmission, where AR token compression, receiver-side prediction, and wireless link adaptation are jointly coordinated under time-varying channels and average symbol-overhead constraints.

\subsection{Contributions}
Existing TokenCom and generative semantic communication studies have mainly focused on exploiting token-level context for recovering missing, corrupted, or untransmitted tokens, and thus reducing communication overhead and latency~\cite{qiao2025todma,qiao2025tokencom,men2026video,ToDMA}. However, the adaptation aspect remains much less explored. 
In particular, a principled digital TokenCom framework that jointly integrates AR token compression, receiver-side token prediction, modulation and coding schemes (MCSs) selection, and long-term symbol-overhead adaptation under time-varying wireless channels is still lacking.

Motivated by this gap, we propose Ada-TokenCom, a rate-adaptive token communication framework for low-rate digital semantic communication. 
The key idea is to jointly exploit AR token compression and receiver-side token prediction. 
Specifically, each token sequence is partitioned into transmitted head tokens and locally generated tail tokens. 
The head tokens are compressed by arithmetic coding (AC) using AR-estimated conditional probabilities and transmitted through a digital wireless link. 
The tail tokens are then predicted at the receiver by the same AR model conditioned on the recovered head tokens. 
This mixed compression-and-generation mechanism enables a controllable rate and quality trade-off by replacing part of explicit token transmission with receiver-side generative computation. 


The main contributions of this work are summarized as follows:
\begin{itemize}[topsep=0.2em, itemsep=0.15em, parsep=0pt, partopsep=0pt]
    \item We propose a rate-adaptive digital TokenCom framework that integrates decoder-only Transformer-based AR probability modeling with AC for bandwidth-efficient token-level compression, which provides a unified digital framework for efficient token compression and transmission.

    \item We design a mixed reconstruction-and-generation mechanism that partitions each token sequence into transmitted head tokens and locally generated tail tokens. 
    This mechanism provides an adaptive rate-quality trade-off by exploiting receiver-side generative priors.

    \item 
    We develop a Lyapunov-based cross-layer optimization algorithm that jointly selects the token truncation length and MCS under time-varying wireless channels and long-term symbol-overhead constraints, thereby opportunistically exploiting favorable channel conditions to improve semantic quality while satisfying the prescribed long-term CBR budget.
\end{itemize}

\textbf{Notations:} 
Bold lowercase, bold uppercase, and normal-face letters denote vectors, matrices, and scalars, respectively. 
$[N]\triangleq\{1,2,\ldots,N\}$. 
For a vector $\mathbf{v}$, $\mathbf{v}^{\mathrm{T}}$,  $[\mathbf{v}]_i$, and $[\mathbf{v}]_{i:j}$ denote its transpose, $i$-th element, and subvector from the $i$-th to the $j$-th element, respectively. 
For a matrix $\mathbf{A}$, $[\mathbf{A}]_{:,j}$ and $[\mathbf{A}]_{:,i:j}$ denote its $j$-th column and the submatrix formed by its $i$-th to $j$-th columns, respectively. $\mathbb{E}[\cdot]$ denotes the expectation operator, and $\mathbb{E}[\cdot\mid\cdot]$ denotes conditional expectation.

\vspace{-2mm}

\section{\color{black}TokenCom with AR and AC Source Coding}
In this section, we present the main components of adaptive token-level source coding in the proposed Ada-TokenCom framework.

\subsection{Token Representation and AR Modeling}
Consider a typical token communication system, where high-dimensional signals such as images or video frames are first split (e.g., into patches) and then discretized by a pretrained tokenizer into sequences of tokens that convey semantic and structural information. Specifically, taking an image $\mathbf{X}\in \mathbb{R}^{H\times W\times C}$ as an example, tokenization transforms $\mathbf{X}$ into $N$ token embeddings, producing a sequence of token indices, denoted as
\begin{equation}
	\mathbf{s} = [s_1, s_2, \dots, s_N]^{\mathrm{T}} = \tau(\mathbf{X}),
\end{equation}
where $\tau(\cdot)$ denotes the tokenization process. Each token index $s_n \in [V]$ points to a codeword in the token codebook $\mathbf{D} \in \mathbb{R}^{V \times d}$, with $V$ being the vocabulary size and $d$ the embedding dimension. 
For notational simplicity, we refer to each discrete token index simply as a ``token'' in the following.

Directly modeling the joint distribution of the token sequence $\mathbf{s}$ conditioned on $\mathbf{s}^c$ is intractable due to its high dimensionality.
Here, $\mathbf{s}^c$ denotes conditioning tokens that provide compact semantic side information, such as textual captions, contextual priors, or class labels. Such side information enables multimodal TokenCom by conveying coarse source semantics. For example, the class label token can represent the semantic concept of a ``Golden Retriever''.
An AR framework makes this tractable by factorizing the joint distribution as
\begin{equation}\label{eq:pmodel}
p(\mathbf{s} \mid \mathbf{s}^c) = \prod_{n=1}^{N} p(s_n \mid [\mathbf{s}]_{1:n-1}, \mathbf{s}^c).
\end{equation}
Each one-step conditional $p(s_n \mid [\mathbf{s}]_{1:n-1}, \mathbf{s}^c)$ can be approximated by a pretrained AR language or multimodal model. Typically, a predefined token ordering, e.g., raster-scan order, is used during training and inference. We adopt the raster-scan order as the default token ordering, although the proposed framework is not limited to raster-scan order and can accommodate other predefined token orderings.

According to the probabilistic model in (\ref{eq:pmodel}), the self-information of a message $\mathbf{s}$ conditioned on $\mathbf{s}^c$ is given by
\begin{equation}\label{eq:SI}
\begin{aligned}
I(\mathbf{s} \mid \mathbf{s}^c)
&= \sum_{n=1}^{N} I(s_n \mid [\mathbf{s}]_{1:n-1}, \mathbf{s}^c) \\
&= - \sum_{n=1}^{N} \log p\!\left(s_n \mid [\mathbf{s}]_{1:n-1}, \mathbf{s}^c \right),
\end{aligned}
\end{equation}
where $I(s_n \mid [\mathbf{s}]_{1:n-1},\mathbf{s}^c) = -\log p\!\left(s_n \mid [\mathbf{s}]_{1:n-1},\mathbf{s}^c \right)$ denotes the self-information of the $n$-th token, $n \in [N]$. This quantity represents the minimum number of bits required to encode the token under the given probabilistic model.
To obtain such a model and examine how 
$I\big(s_n\mid[\mathbf{s}]_{1:n-1},\mathbf{s}^c\big)$ 
varies with the token position $n$, we employ pretrained AR image generation models. During the encoding process,
images are first tokenized into discrete latent tokens, which are then processed by a pretrained Transformer in raster-scan order to produce AR probability distributions. 
These probabilities allow us to estimate the conditional self-information at each token position.
The conditional self-information generally decreases along the generation order, indicating that earlier tokens contribute more information than later tokens. The resulting profile characterizes the self-information associated with each token position, and equivalently the number of bits required to encode the token conditioned on the preceding tokens.

In addition to the realized self-information of the ground-truth token, the AR predictive distribution also provides a measure of model uncertainty before observing the token value. 
The corresponding conditional predictive entropy is defined as
{\setlength{\abovedisplayskip}{4pt}
 \setlength{\belowdisplayskip}{8pt}
 \setlength{\abovedisplayshortskip}{3pt}
\begin{equation}
\begin{aligned}
H_n
=
-\sum_{v=1}^{V}
&p(s_n=v \mid [\mathbf{s}]_{1:n-1}, \mathbf{s}^c) \\
&\times
\log_2 p(s_n=v \mid [\mathbf{s}]_{1:n-1}, \mathbf{s}^c).
\end{aligned}
\label{eq:cond_entropy}
\end{equation}}
Unlike self-information, $H_n$ quantifies the uncertainty of the predictive distribution. Lower entropy indicates that tail tokens can be more reliably generated from the available context. 
This predictive uncertainty provides an interpretation for tail-token truncation. When the prefix tokens already provide sufficient semantic and structural context, the conditional distribution of later tokens can become more concentrated, leading to lower predictive entropy. In such cases, even if the tail tokens are not explicitly transmitted, the AR model may still complete them in a way that is consistent with the main semantics of the original image. 

These observations motivate the use of learned AR probabilities for source coding: if the probability distributions of future tokens can be predicted from previously decoded tokens and side information, then the expected coding rate can be reduced accordingly.

\begin{figure*}[!t]
	\centering
	\subfloat{\includegraphics[width=1\textwidth]{ 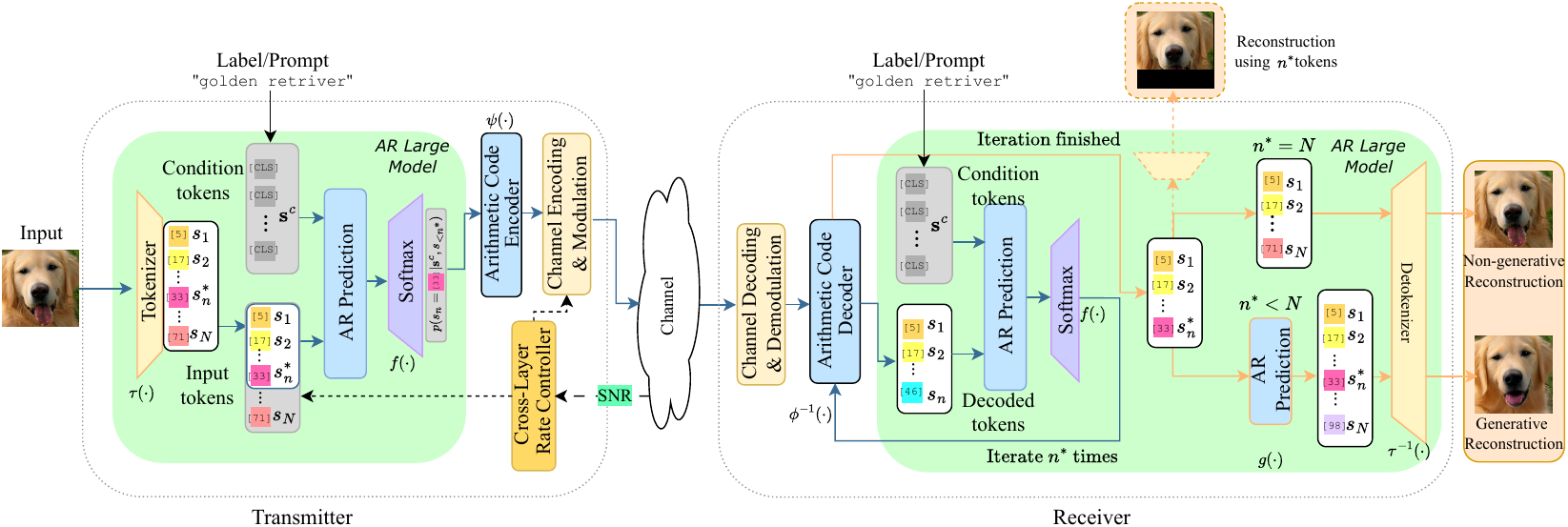}}
	\caption{Block diagram of the proposed {\color{black}Ada-TokenCom} framework. 
    }
	\label{Fig.1}
    \vspace{-3mm}
\end{figure*}

\subsection{\color{black}Arithmetic Coding with Large AR Model-Based Priors}

AR modeling and AC are naturally compatible, since both operate sequentially on prefix-conditioned probability distributions. 
In Ada-TokenCom, the AC encodes the token sequence by recursively refining a subinterval of $[0,1)$, using the AR model's estimation of the factorized probability \eqref{eq:pmodel}.
From an information-theoretic perspective, a more accurate predictive distribution produced by the AR model yields a smaller expected code length. In particular, ignoring the negligible implementation overhead of practical coders, the resulting codelength for AC is close to the self-information described in (\ref{eq:SI}). 

Let $f(\cdot)$ denote the pretrained AR model that outputs the predictive distribution over the token vocabulary. We define $[\mathbf{s}]_{1:0}\triangleq\emptyset$ as the empty prefix. For $n\in [n^*]$, the predictive distribution for the $n$-th token is given by
\begin{equation}
\mathbf{p}^n = f([\mathbf{s}]_{1:n-1},\mathbf{s}^c),
\label{eq:prob}
\end{equation}
where $\mathbf{p}^n\in\mathbb{R}^{V}$ and $[\mathbf{s}]_{1:0}\triangleq\emptyset$ denotes the empty prefix. Accordingly, for each candidate token $v\in[V]$, we have
\begin{equation}
[\mathbf{p}^n]_v
=
p(s_n=v\mid[\mathbf{s}]_{1:n-1},\mathbf{s}^c).
\end{equation}
Collecting the predictive distributions over all positions yields
\begin{equation}
\mathbf{\Pi}
=
\left[
\mathbf{p}^1, \mathbf{p}^2, \dots, \mathbf{p}^{n^*}
\right]
\in \mathbb{R}^{V\times n^*}.
\label{eq:prob_sequence}
\end{equation}
According to~\cite{Witten1987}, AC starts from the initial interval $\mathcal{J}_0=[0,1)$. For encoding the $n$-th token in the sequence, the current interval is $\mathcal{J}_{n-1}=[l_{n-1},u_{n-1})$. Define the cumulative distribution associated with $\mathbf{p}^n$ as
\begin{equation}
F_n(v)=\sum_{j=1}^{v}[\mathbf{p}^n]_j,
\qquad
F_n(0)=0.
\end{equation}
Then, for any candidate token value $v\in[V]$, the corresponding subinterval is given by
\begin{equation}\label{eq:interv}
\begin{aligned}
\mathcal{J}_n(v) =& \bigg[l_{n-1} + (u_{n-1}-l_{n-1})F_n(v-1),\\
& ~~~~~~~~~l_{n-1} + (u_{n-1}-l_{n-1})F_n(v)\bigg).
\end{aligned}
\end{equation}
Given the actual token $s_n$, the encoder selects the corresponding subinterval
\begin{equation}
\mathcal{J}_n = \mathcal{J}_n(v|v=s_n),
\qquad n=1,2,\dots,n^*.
\end{equation}
By recursively refining the interval over all $n^*$ tokens in the sequence, AC maps the sequence $\mathbf{s}$ to a unique final interval $\mathcal{J}_{n^*}$. Any representative value $z \in \mathcal{J}_{n^*}$ is first selected as the code value associated with the final interval. The value $z$ is then converted into a binary fractional representation, yielding the compressed bitstream $B_{n^*}$ carrying the predictive conditional probability vectors of the first $n^*$ token positions. The AC encoding procedure can be written as
\begin{equation}
B_{n^*}
=
\psi\!\left(\mathbf{s}^c, [{\mathbf{s}}]_{1:n^*},
[\mathbf{\Pi}]_{:,1:n^*}
\right),
\label{eq:bitstream_generation}
\end{equation}
where $\psi(\cdot)$ denotes the AC encoder driven by the AR-model probability sequence. At the decoder, the same conditional distributions are reproduced from $\mathbf{s}^c$ and the previously decoded prefix using the pre-trained large AR model, so that the token sequence can be recovered in a lossless manner.

\subsection{\color{black}Mixed Reconstruction/Generation for Compression Rate Adaptation}

Given the token sequence $\mathbf{s}_{n^*}$ produced by the tokenizer, AC can losslessly compress the transmitted tokens according to the conditional probabilities provided by the AR model.
This observation further motivates a mixed reconstruction/generation strategy for rate adaptation in Ada-TokenCom, i.e., instead of AC-encoding the entire token sequence, the transmitter encodes only the more informative head tokens, while the remaining tail tokens are not transmitted and are directly predicted by the AR model at the receiver. In this way, both the shared side information $\mathbf{s}^c$ and the decoded token prefix serve as conditions to guide the prediction of the remaining tail tokens.

\subsubsection{\color{black}Reconstruction with AC-based Head Token Decoding}

The receiver employs an identical large AR model to reproduce the same conditional probability distributions as those at the transmitter. Before decoding, the receiver initializes the interval $\mathcal{J}_0=[0,1)$ in the same manner as the transmitter. Conceptually, the received bitstream $\hat{B}_{n^*}$ specifies the code value $ \hat{z}$ contained in the final coding interval. Decoding is then performed sequentially from the first token to the $n^*$-th transmitted token.

For $n=1,\ldots,n^*$, the receiver reproduces the predictive distribution $\hat{\mathbf p}^n$ according to~\eqref{eq:prob}. The predictive distribution partitions the current interval $\hat{\mathcal J}_{n-1}$ into $V$ subintervals corresponding to candidate token indices $v\in[V]$. The decoder then determines $\hat{s}_n$ as the unique token whose associated subinterval contains the code value $\hat z$. Accordingly, the decoding rule is given by
\begin{equation}
\hat{s}_n = v
\quad \text{if and only if} \quad
\hat{z} \in \hat{\mathcal{J}}_n(v),
\qquad v\in[V],\; n\in[n^*],
\label{eq:ac_decoding_rule}
\end{equation}
where $\hat{\mathcal{J}}_n(v)$ is the subinterval induced by the predictive distribution at step $n$ from the current interval $\hat{\mathcal{J}}_{n-1}$ according to \eqref{eq:interv}. After $\hat{s}_n$ is determined, the interval is updated as $\hat{\mathcal{J}}_n=\hat{\mathcal{J}}_n(v|v=\hat{s}_n)$, and the procedure continues recursively until the first $n^*$ tokens have been recovered.
For notational simplicity, we denote by $\psi^{-1}(\cdot)$ the arithmetic-decoding operator that recovers a token according to the decoded conditional probability vector. Then, the $n$-th decoded token can be written as
\begin{equation}
\hat{s}_n
=
\psi^{-1}\!\left([\hat{\mathbf{\Pi}}]_{:,1:n},\hat{B}_{n}\right),
\qquad n\in[n^*].
\end{equation}

\subsubsection{\color{black}Generation with Large AR Model-based Tail Token Prediction}

After the first $n^*$ tokens have been recovered, the receiver uses the large AR model to autoregressively predict the remaining $(N-n^*)$ tail tokens. Since the shared side information $\mathbf{s}^c$ is available at both the transmitter and the receiver, and the decoded prefix $[\hat{\mathbf{s}}]_{1:n^*}$ has already been recovered at the receiver, these two sources of context jointly guide the prediction of the missing tokens. We define $g(\cdot)$ as the token prediction function of the large AR model. 
The overall token reconstruction process can be expressed as
\begin{equation}
\hat{s}_n =
\begin{cases}
\psi^{-1}\!\left([\hat{\mathbf{\Pi}}]_{:,1:n},\hat{B}_{n}\right), & n\in[n^*], \\[2mm]
g(\mathbf{s}^c,[\hat{\mathbf{s}}]_{1:n-1}), & n^* < n \leq N.
\end{cases}
\label{eq:recon}
\end{equation}
Therefore, the first $n^*$ head tokens are recovered exactly in the token domain via AC decoding, whereas the remaining tail tokens are completed by conditional AR prediction.
The reconstructed signal $\hat{\mathbf{X}}$ is finally obtained as
\begin{equation}
\hat{\mathbf{X}} = \tau^{-1}([\hat{s}_1,\hat{s}_2,\cdots,\hat{s}_N]^{\mathrm{T}}),
\end{equation}
where $\tau^{-1}(\cdot)$ denotes the detokenization operator. 
This mechanism enables straightforward compression-rate adaptation by adjusting the transmitted prefix token length: under tighter rate constraints, a smaller $n^*$ is transmitted and more tail tokens are generated at the receiver; under looser rate constraints, a larger transmitted prefix enables more faithful reconstruction. 
In this way, the proposed TokenCom framework combines exact AC-based recovery of the transmitted head tokens with large-AR-model-based completion of the untransmitted tail tokens, thereby achieving flexible rate adaptation while preserving semantic consistency. 

\section{Cross-Layer Rate Adaptation for Wireless TokenCom} \label{sec:lyap}

In wireless TokenCom, both the packet length and the selected channel protection level directly affect semantic reliability under limited communication resources. In the proposed Ada-TokenCom framework, the retained prefix length $n^*$ determines the source rate, while the selected MCS determines the channel protection level. This coupling naturally motivates a cross-layer rate adaptation mechanism that jointly controls source truncation and wireless transmission. This design choice cleanly separates source compression from channel protection, thereby maintaining compatibility with the separation-based design of the protocol stack in the existing wireless networks.

Specifically, the transmitter encodes only the first $n^*$ tokens, where $n^* \in [N]$, into a bitstream via AC, while the receiver reconstructs the remaining tail tokens using the large AR model conditioned on the decoded prefix and the shared side information $\mathbf{s}^c$. For a single packet, adjusting $n^*$ directly changes the packet length, which in turn affects the packet error rate (PER), increases the risk of packet retransmissions, and impacts the achievable semantic fidelity under limited channel resources. To exploit this coupling over time-varying wireless channels, we jointly optimize the retained prefix length and the channel transmission strategy through a Lyapunov-based cross-layer adapter~\cite{Neely2010}.

\subsection{Wireless Communication Model}

We consider a point-to-point wireless communication system and model the wireless link as a quasi-static block-fading Rayleigh channel. The transmitter applies channel encoding and $M$-ary quadrature amplitude modulation ($M$-QAM) to the compressed bitstream $\mathbf{b}$, producing a normalized baseband symbol vector for transmission over the wireless channel. The channel coefficient follows a circularly symmetric complex Gaussian distribution with zero mean and unit variance, and remains constant within one transmission block, while the additive noise is modeled as circularly symmetric complex Gaussian noise.
Assuming perfect channel state information at the receiver, zero-forcing equalization is adopted before soft demodulation and channel decoding. The receiver then performs cyclic redundancy check (CRC) verification to determine whether the packet is successfully received.

\subsection{PER Approximation for Online Link Adaptation}

In each time slot, the transmitter jointly selects the retained prefix length $n^*$ and a MCS $m \in \mathcal{M}$ according to the cross-layer optimization decision. Let $\ell(n^*)$ denote the number of source bits generated after AC encoding of the first $n^*$ tokens. After applying channel coding and appending protocol overhead such as CRC bits, the total packet length is denoted by $L$. Since the packet error rate depends on both the packet length and the selected MCS, we characterize the error behavior for each MCS $m$ by a pair of fitting coefficients, which are obtained offline as described below.

Although the wireless link considered in our system model follows a quasi-static Rayleigh fading channel, directly evaluating the average PER for every candidate transmission configuration is computationally inconvenient for online adaptation. To enable efficient online transmission adaptation, we therefore adopt the waterfall-threshold approximation in \cite{4558591}. Specifically, for a given MCS $m$ and packet length $L$, the packet error behavior over an additive white Gaussian noise (AWGN) channel is summarized by a waterfall threshold $\gamma_\mathrm{waterfall}^{m}(L)$. This threshold is then used to approximate the average PER over the quasi-static Rayleigh fading channel as
\begin{equation}
\mathrm{PER}^m(L,\bar{\gamma})
\approx
1-\exp\!\left(
-\frac{\gamma_\mathrm{waterfall}^{m}(L)}{\bar{\gamma}}
\right),
\label{eq:per_threshold}
\end{equation}
where $\bar{\gamma}$ denotes the average received SNR~\cite{4558591}.

Following \cite{6514951}, the waterfall threshold for a fixed MCS $m$ can be further approximated as a linear function of the packet length in the logarithmic domain, i.e.,
\begin{equation}
\gamma_\mathrm{waterfall}^{m}(L)\approx \alpha_m \ln L+\beta_m,
\label{eq:gammaw_fit}
\end{equation}
where $\alpha_m$ and $\beta_m$ are coefficients obtained by curve fitting offline. Substituting \eqref{eq:gammaw_fit} into \eqref{eq:per_threshold} yields
\begin{equation}
\mathrm{PER}^m(L,\bar{\gamma})
\approx
1-\exp\!\left(
-\frac{\alpha_m \ln L+\beta_m}{\bar{\gamma}}
\right).
\label{eq:per_final}
\end{equation}

\subsection{Lyapunov-Based Joint Prefix-Length and MCS Adaptation}\label{subsec:lyap}
In the proposed Ada-TokenCom framework, the retained prefix length $n^*$ has a dual effect: it determines the packet length and hence the transmission reliability, while also adapting how many tail tokens are predicted by conditional AR prediction at the receiver. Therefore, the choice of $n^*$ directly affects the trade-off between communication efficiency and semantic reconstruction fidelity.

It is worth noting that reducing the $n^*$ value mainly affects the communication resource consumption, without affecting the receiver-side inference complexity. This is because the receiver still traverses the full token sequence: the transmitted head tokens are reconstructed sequentially via AC-assisted decoding, while the remaining tail tokens are generated by AR predictions with the same large model. Hence, we focus on the dominant communication--semantics trade-off and jointly adapt the retained prefix length $n^*$ and the MCSs.

Since the wireless channel varies over time, the desirable prefix length--MCS pair may also vary from slot to slot. In favorable channel conditions, transmitting a longer prefix can improve semantic fidelity with acceptable transmission cost; while in poor channel states, a shorter prefix together with stronger channel protection may be preferable. Therefore, instead of imposing a rigid per-slot resource constraint, we constrain only the long-term average channel-symbol consumption. This allows the adapter to exploit favorable channel states opportunistically while respecting the overall average communication resource budget. Under this stochastic long-term formulation, Lyapunov optimization provides an effective online adaptive method without requiring knowledge of future channel realizations.

Assume that one image is sent per time slot. In $t$-th time slot, the transmitter selects a retained prefix length $n_t^*$ and an MCS $m_t$. 
Define the decision variable as
$
d_t \in\mathcal D,
$
where
$
\mathcal D \triangleq
\left\{
(n_t^*,m_t)\,\middle|\,n_t^*\in\mathcal N,\;m_t\in\mathcal M
\right\}
$
is the set of all feasible prefix-length--MCS pairs, $\mathcal N$ denotes the candidate prefix-length set, and $\mathcal M$ denotes the candidate MCS set.
For a given decision $d_t$, let $L_t$ and $C(d_t)$ denote the packet length and the number of channel symbols for one transmission attempt, respectively. According to \eqref{eq:per_final}, the PER is approximated as
\begin{equation}
P_t(d_t,\bar{\gamma}_t) \triangleq \mathrm{PER}^{m_t}(L_t,\bar{\gamma}_t),
\end{equation}
where $\bar{\gamma}_t$ is the average received SNR in $t$-th slot. 
To account for transmission failures, type-I automatic repeat request (ARQ) is employed, where a packet is retransmitted in full whenever decoding fails~\cite{linarq}. Under the standard geometric approximation, the expected number of transmission attempts required for successful delivery is $1/(1-P_t(d_t,\bar{\gamma}_t))$. Hence, the expected channel-symbol consumption can be considered as 
\begin{equation}
\bar{C}_t(d_t,\bar{\gamma}_t)
=
\frac{C(d_t)}{1-P_t(d_t,\bar{\gamma}_t)}.
\label{eq:expected_symbol_consumption}
\end{equation}
Let $A_t(n_t^*)$ denote the semantic quality achieved in $t$-th slot, e.g., peak signal-to-noise ratio (PSNR) for pixel-level fidelity, learned perceptual image patch similarity (LPIPS) \cite{zhang2018unreasonable} for perceptual similarity, contrastive language--image pre-training (CLIP) similarity \cite{pmlr-v139-radford21a} for semantic alignment, etc. Here, a CLIP-based semantic similarity score is adopted. The long-term objective is to maximize the time-average semantic quality subject to an average channel-symbol budget
\begin{equation}
\begin{aligned}
\max_{\{d_t\in\mathcal{D}\}} \quad
& \lim_{T\to\infty}\frac{1}{T}\sum_{t=1}^{T}
\mathbb{E}\!\left[A_t(n_t^*)\right] \\
\text{s.t.}\quad
& \lim_{T\to\infty}\frac{1}{T}\sum_{t=1}^{T}
\mathbb{E}\!\left[\bar{C}_t(d_t,\bar{\gamma}_t)\right]
\leq C_{\mathrm{th}},
\end{aligned}
\label{eq:cross_layer_problem}
\end{equation}
where $C_{\mathrm{th}}$ is the allowable average number of channel symbols per slot.
Following the classical Lyapunov optimization framework in~\cite{Neely2010}, we introduce a virtual queue $\{Z_t\}$ with initial value $Z_0=0$ to enforce the long-term channel-symbol budget constraint by tracking the accumulated budget violation. The queue evolves according to
\begin{equation}
Z_{t+1}
=
\max\!\left\{
Z_t+\bar C_t(d_t,\bar\gamma_t)-C_{\rm th},\,0
\right\},
\quad t=0,1,2,\ldots,
\label{eq:virtual_queue}
\end{equation}
where the queue accumulates the excess expected channel-symbol consumption relative to the prescribed budget $C_{\rm th}$. Intuitively, aggressive transmission decisions increase the queue backlog and consequently impose a larger penalty on future resource-intensive actions.
According to~\cite{Neely2010}, we define the Lyapunov function and the one-slot conditional drift as
$
\mathcal L(Z_t)\triangleq \frac12 Z_t^2$ and
$
\Delta(Z_t)
\triangleq
\mathbb E\!\left[
\mathcal L(Z_{t+1})-\mathcal L(Z_t)
\mid Z_t
\right]
$, respectively.
The drift can be upper-bounded as
\begin{equation}
\begin{aligned}
\Delta(Z_t)
&=
\mathbb E\!\left[
\frac12\!\left(Z_{t+1}^2-Z_t^2\right)
\Bigm| Z_t
\right]
\\
&\le
\mathbb E\!\left[
\frac12
\left(
Z_t+\bar C_t(d_t,\bar\gamma_t)-C_{\rm th}
\right)^2
-\frac12 Z_t^2
\Bigm| Z_t
\right]
\\
&\le
\kappa
+
Z_t\,
\mathbb E\!\left[
\bar C_t(d_t,\bar\gamma_t)-C_{\rm th}
\mid Z_t
\right],
\end{aligned}
\label{eq:drift_bound}
\end{equation}
where $\kappa$ is a constant independent of the adaptation decision.
To jointly maximize semantic quality while stabilizing the virtual queue, the drift-plus-penalty principle and upper-bound the following objective:
\begin{equation}
\begin{aligned}
&\Delta(Z_t)
-
\eta
\mathbb E\!\left[
A_t(n_t^*)
\mid Z_t
\right]
\\
&\le
\kappa
+
Z_t
\mathbb E\!\left[
\bar C_t(d_t,\bar\gamma_t)-C_{\rm th}
\mid Z_t
\right]
-
\eta
\mathbb E\!\left[
A_t(n_t^*)
\mid Z_t
\right],
\end{aligned}
\label{eq:drift_minus_quality_bound}
\end{equation}
where $\eta>0$ is an adaptation parameter balancing semantic quality and long-term channel-symbol consumption.
Since $\kappa$ and $-Z_tC_{\rm th}$ are independent of the adaptation decision $d_t$, minimizing the right-hand side of \eqref{eq:drift_minus_quality_bound} is equivalent to solving
\begin{equation}
\begin{aligned}
d_t^*
&=
\arg\min_{d_t\in\mathcal D}
\Bigg\{
Z_t
\mathbb E\!\left[
\bar C_t(d_t,\bar\gamma_t)
\mid Z_t
\right]
-
\eta
\mathbb E\!\left[
A_t(n_t^*)
\mid Z_t
\right]
\Bigg\}
\\
&=
\arg\min_{d_t\in\mathcal D}
\left\{
Z_t\bar C_t(d_t,\bar\gamma_t)
-
\eta A_t(n_t^*)
\right\}.
\label{eq:per_slot_policy}
\end{aligned}
\end{equation}
Therefore, in each slot, the transmitter searches over the feasible set of prefix length--MCS pairs and selects the configuration that minimizes \eqref{eq:per_slot_policy}. In this way, the adapter explicitly balances communication efficiency and semantic fidelity under time-varying wireless conditions. The overall procedure is summarized in Algorithm~\ref{alg:lyapunov_joint_control}.

\begin{algorithm}[!t]
\small
\caption{\small The Proposed Online Joint Prefix-Length and MCS Adaptation for Ada-TokenCom}
\label{alg:lyapunov_joint_control}
\begin{algorithmic}[1]
\Require Candidate prefix-length set $\mathcal{N}$, candidate MCS set $\mathcal{M}$, adaptation parameter $\eta$, average symbol budget $C_{\mathrm{th}}$, average received SNR $\bar{\gamma}_t$
\State Initialize $Z_0=0$.
\For{$t=1,2,\ldots$}
    \For{each $d_t\in\mathcal{D}$}
        \State Determine the corresponding packet length $L_t$ and one-shot channel-symbol consumption $C(d_t)$.
        \State Compute $P_t(d_t,\bar{\gamma}_t)$ according to \eqref{eq:per_final}.
        \State Compute $\bar{C}_t(d_t,\bar{\gamma}_t)$ according to \eqref{eq:expected_symbol_consumption}.
        \State Evaluate the minimization objective in \eqref{eq:per_slot_policy}.
    \EndFor
    \State Obtain $d_t^*$ according to \eqref{eq:per_slot_policy}.
    \State Update $Z_{t+1}$ according to \eqref{eq:virtual_queue}.
\EndFor
\end{algorithmic}
\end{algorithm}

\subsection{Complexity Analysis of Online Adaptation}
For each candidate $d_t\in\mathcal{D}$, the evaluation of \eqref{eq:per_final}, \eqref{eq:expected_symbol_consumption}, and \eqref{eq:per_slot_policy} requires only constant-time operations, while the mapping from prefix length $n$ to the corresponding bit length $\mathbf{b}_n$ can be implemented via lookup. Therefore, the per-slot complexity of the proposed online adapter is
$
\mathcal{O}(|\mathcal{D}|)=\mathcal{O}(|\mathcal{N}||\mathcal{M}|).
$
The virtual queue update in \eqref{eq:virtual_queue} has complexity $\mathcal{O}(1)$.

\section{Simulation Results}

\subsection{Experimental Setup}

To evaluate the decoding performance of the proposed Ada-TokenCom framework, we adopt the tokenizer and generator architecture provided by LlamaGen \cite{sun2024autoregressive}, a GPT-style large image generation model, where both the image tokenizer and the AR transformer are pre-trained following \cite{sun2024autoregressive}. In the main experiments, we consider the class-conditioned image generation setting, i.e., class-to-image (c2i). Specifically, a vector quantized generative adversarial networks (VQGAN)-based tokenizer trained on ImageNet-1K is combined with the corresponding LlamaGen-L AR model. For each image, the tokenizer produces a $16\times16$ discrete latent representation grid, each latent token corresponds to a spatial region of the original image. The tokenizer then flattens the grid, and maps it into a sequence of 256 visual tokens using a codebook with size \(V=16384\). The image resolution is $256\times 256$ unless otherwise specified. 
More generally, the semantic condition is not limited to a class label and may also take other forms of high-level semantic guidance, such as text descriptions. In this paper, the class-conditioned setting is used for the main quantitative evaluation, whereas other conditional forms are presented later only for illustrative discussion.
Performance is evaluated on the ImageNet-V2 data set \cite{recht2019imagenet}, which shares the same label space as ImageNet-1K but is not used for training, thereby assessing out-of-sample generalization. All experiments are conducted on a NVIDIA 5090 GPU.

\subsection{Self-Information and Entropy Analysis of AR Image Tokens}
We begin by empirically examining the key observation introduced above: the conditional uncertainty of an image token can be reduced when its head tokens and side information are available. Following~\eqref{eq:SI}, we compute self-information in bits at each position of the token sequence $\mathbf{s}$, i.e., 
$I(s_n\mid [\mathbf{s}]_{1:n-1},\mathbf{s}^c)$. 
Specifically, we evaluate two AR models from LlamaGen with different parameter scales, namely LlamaGen-B and LlamaGen-L. 
For comparison, we also include \emph{Taming Transformer}~\cite{Esser2021CVPR}, which follows an identical AR modeling paradigm, and shares an identical codebook size $V=16384$ with LlamaGen-B and LlamaGen-L. 
This comparison allows us to examine whether the observed self-information behavior is consistent across different pretrained models.

\begin{figure}[!t]
    \subfloat[]{\includegraphics[width=0.238\textwidth]{ 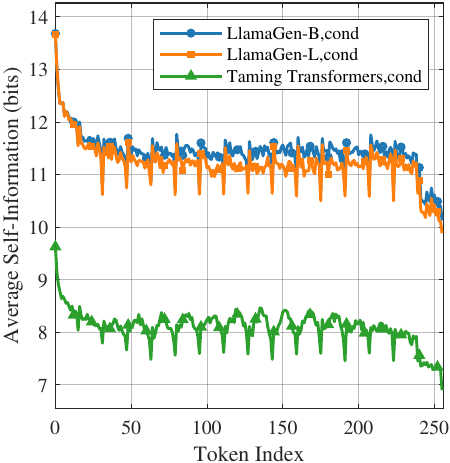}}
    \subfloat[]{\includegraphics[width=0.238\textwidth]{ 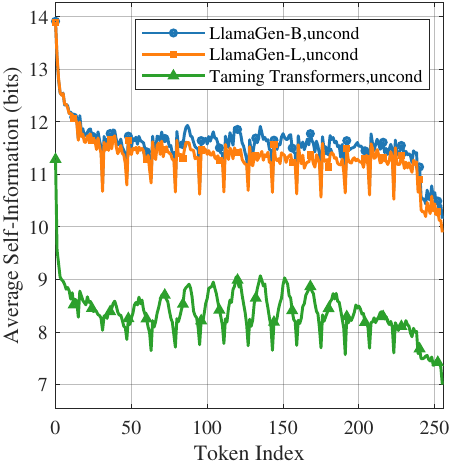}}

    \captionsetup{justification=justified, singlelinecheck=false}

    \caption{Self-information as a function of token position (a) with class label as conditional information (\emph{cond}) and (b) with null label (\emph{uncond}), averaged over 1000 images from ImageNet-V2 datatset, for pretrained AR models with different parameter scales.}
    \label{fig:SI}
    \vspace{-3mm}
\end{figure}

\begin{figure}[!t]
    \centering
    \subfloat{\includegraphics[width=0.48\textwidth]{ 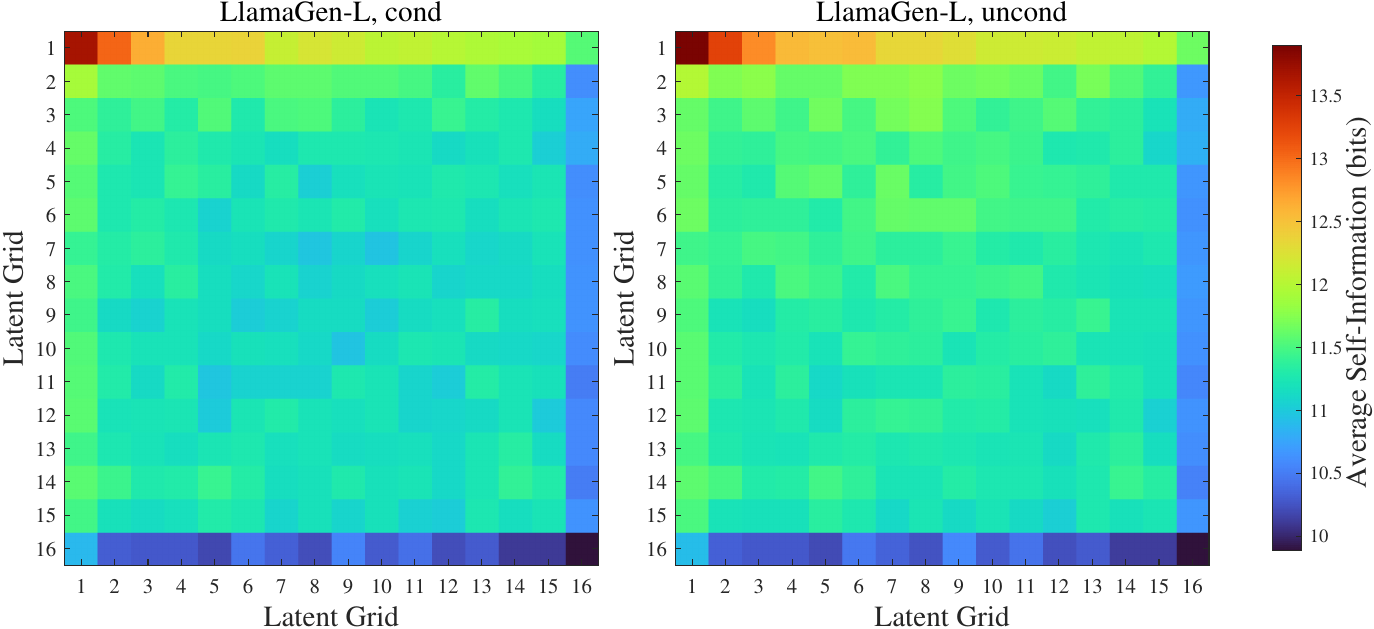}}
    \captionsetup{justification=justified,singlelinecheck=false}
    \caption{Visualization of self-information on a $16\times16$~VQGAN latent grid using LlamaGen-L. Fluctuations predominantly occur at positions associated with discontinuities, which are induced by flattening the 2D latent grid and mapping the grid into 1D token sequence.}
    \label{fig.heatmap}
\end{figure}

We randomly sample 1,000 images from the ImageNet-V2 dataset as inputs and report the averaged results. 
Fig.~\ref{fig:SI} shows the average conditional self-information across token positions with conditional class labels and with null labels (i.e., {\it cond} and {\it uncond}), where null label means that no class label is used for conditional side information. Fig.~\ref{fig.heatmap} visualizes the average self-information for token positions within the VQGAN grid, where the 2D latent grid is flattened row-by-row into a 1D token sequence, and the $i$-th row maps to token positions $16 (i-1) + 1$ through $16i$, where $i\in[16]$.
Self-information is higher for early tokens and decreases as more previous tokens are observed, confirming that preceding tokens provide useful context for subsequent prediction. 
Table~\ref{tab:model_bits} reports the average self-information per token for different pretrained models. 
Larger or stronger models achieve lower average self-information, e.g., 11.42 bits/token for LlamaGen-B, 11.22 bits/token for LlamaGen-L, and 8.09 bits/token for Taming Transformers when class label is available, demonstrating improved probability estimation. For reference, directly representing token indices in binary form requires \(\log_2 V = 14\) bits per token, corresponding to the maximum conditional self-information of a uniform distribution over the codebook.

\begin{figure}[!t]
    \subfloat[]{\includegraphics[width=0.238\textwidth]{ 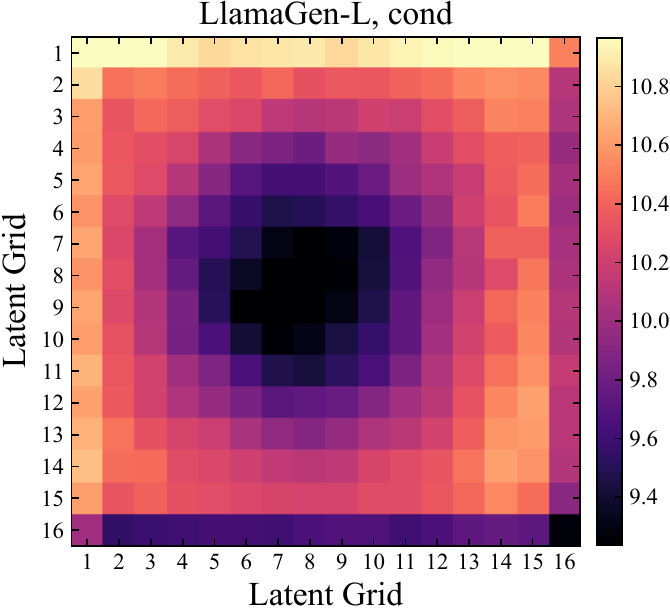}} 
    \subfloat[]{\includegraphics[width=0.238\textwidth]{ 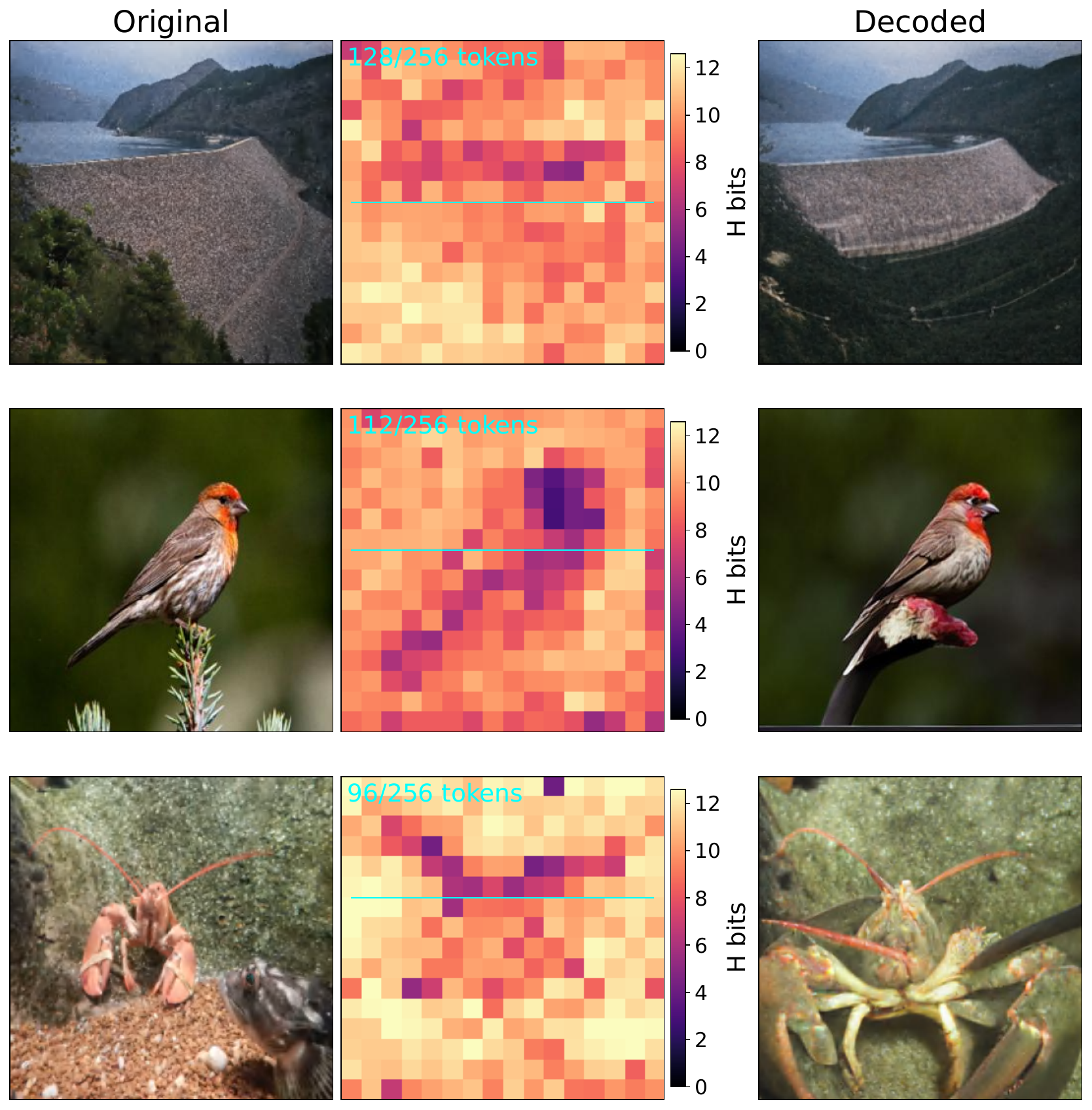}}

    \captionsetup{justification=justified, singlelinecheck=false}

    \caption{Visualization for (a) heatmap of conditional entropy on a $16\times16$~VQGAN latent grid, and (b) samples of their decoded image with specified lengths of head tokens.}
    \label{fig:entropy_heatmap}
\end{figure}
Fig.~\ref{fig:entropy_heatmap}(a) shows the conditional predictive entropy averaged over 1000 ImageNet-V2 images. The entropy is not uniformly distributed across the latent grid. Since the main objects in ImageNet-V2 images are typically located near the image center, the model exhibits a clear low-entropy region in the middle of the latent grid when conditioned on both the class label and the preceding raster-scan tokens. This indicates that, after sufficient prefix information has been observed, subsequent tokens can be predicted more reliably from the available contextual information.
Fig.~\ref{fig:entropy_heatmap}(b) further visualizes decoded examples under different truncation positions. Although the truncation positions vary across images, the decoded images still preserve the main semantic content. This supports that when the class condition and prefix tokens already provide sufficient semantic and structural information, the remaining tail tokens can be plausibly reconstructed by the AR model.
\begin{table}[!t]
\centering

\caption{\centering Average conditional self-information per token under different AR models.}
\label{tab:model_bits}
\small
\setlength{\tabcolsep}{2.8pt}
\begin{tabular}{lcccc}

\toprule
\textbf{Model} & \textbf{Params}\textsuperscript{1} & \makecell{\textbf{Inference}\\ \textbf{Cost}\textsuperscript{2}}
& \makecell{\textbf{Avg. Bits}\\ \textbf{(cond)}}
& \makecell{\textbf{Avg. Bits}\\ \textbf{(uncond)}} \\
\midrule
LlamaGen-B      & 72 M + 111 M   & 52.56 G  & 11.42 & 11.54 \\
LlamaGen-L      & 72 M + 343 M   & 173.57 G & 11.22 & 11.33 \\
Taming Transf.  & 76 M + 1411 M  & 364.00 G & 8.09  & 8.37  \\
\bottomrule
\end{tabular}

\vspace{1mm}
\begin{minipage}{0.95\linewidth}
\footnotesize
\textsuperscript{1} Parameters are reported as tokenizer parameters + AR model parameters.\\
\textsuperscript{2} Inference cost is evaluated for 256 tokens and measured in floating-point operations (FLOPs).
One fused multiply-add is counted as 2 FLOPs.
\end{minipage}
\end{table}

\subsection{Semantic Compression Performance}
In this subsection, the proposed framework is compared with several representative learned image compression baselines, including {\it bmshj2018-Hyperior} \cite{balle2018variational}, {\it Cheng2020-Attention} \cite{Cheng_2020_CVPR} implemented using the CompressAI repository \cite{begaint2020compressai}, {\it HiFiC} \cite{mentzer2020high}, {\it Better Portable Graphics} (BPG), and GenSC scheme powered by diffusion models proposed in~\cite{10599525}, namely {\it Diff. GenSC}, with prompt ``\texttt{A photo of [label\_name]}''. For token-based reference schemes, we use both the {\it LlamaGen-L} and {\it Taming Transformers} as the backbone models of proposed {\it Ada-TokenCom}. We further compare the performance under {\it cond} and {\it uncond} settings, and compare our framework with {\it TokenCom} \cite{qiao2025tokencom} in which each token is directly represented using $\log_2 V = 14$ bits without AR completion.
For Ada-TokenCom, different compression rates are realized by truncating the transmitted token sequence to different lengths, and the resulting rate is measured in bits per pixel (bpp).
Reconstruction quality is evaluated on a randomly selected subset of 1000 validation images from ImageNet-V2, covering diverse classes. We adopt complementary metrics: PSNR, LPIPS, CLIP similarity, and Fr\'echet inception distance (FID) \cite{NIPS2017_8a1d6947} for distributional realism. Since FID is computed on the selected subset rather than on the full benchmark, its absolute values are not directly comparable to those reported in \cite{sun2024autoregressive}.

\begin{figure*}[!t]
    \centering
    \captionsetup{font={footnotesize}, singlelinecheck=off, labelsep=period}
    \subfloat{\includegraphics[width=1\textwidth]{ 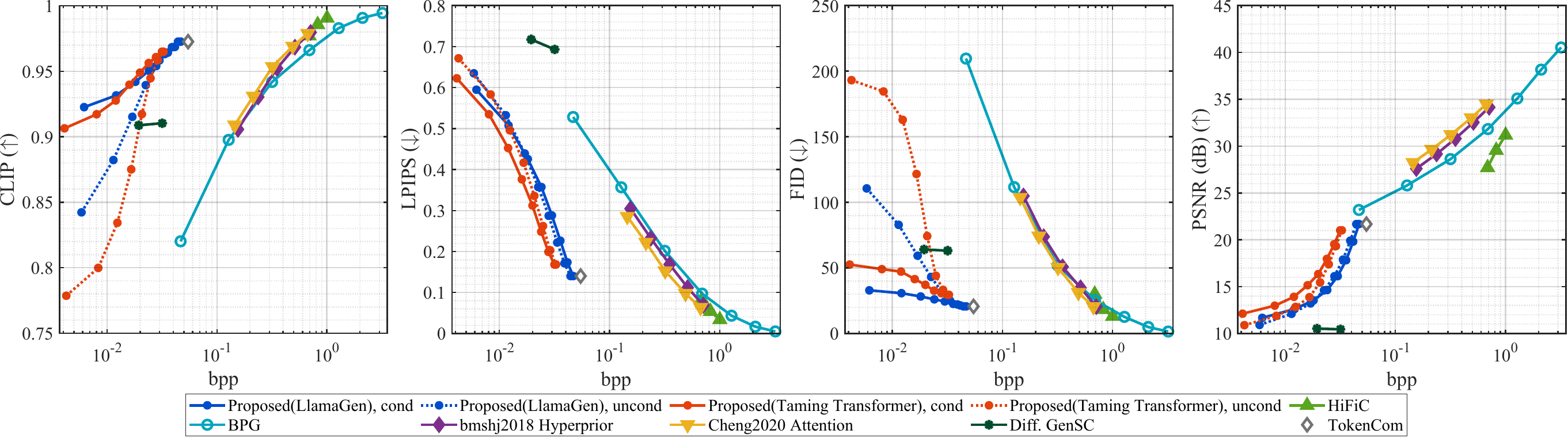}}

    \caption{Performance comparison between the proposed 
    Ada-TokenCom framework and benchmark image compression 
    methods on the ImageNet-V2 dataset.}
    \label{Fig.2}
\end{figure*}
\begin{figure*}[t]
    \centering
    \subfloat{\includegraphics[width=0.95\textwidth]{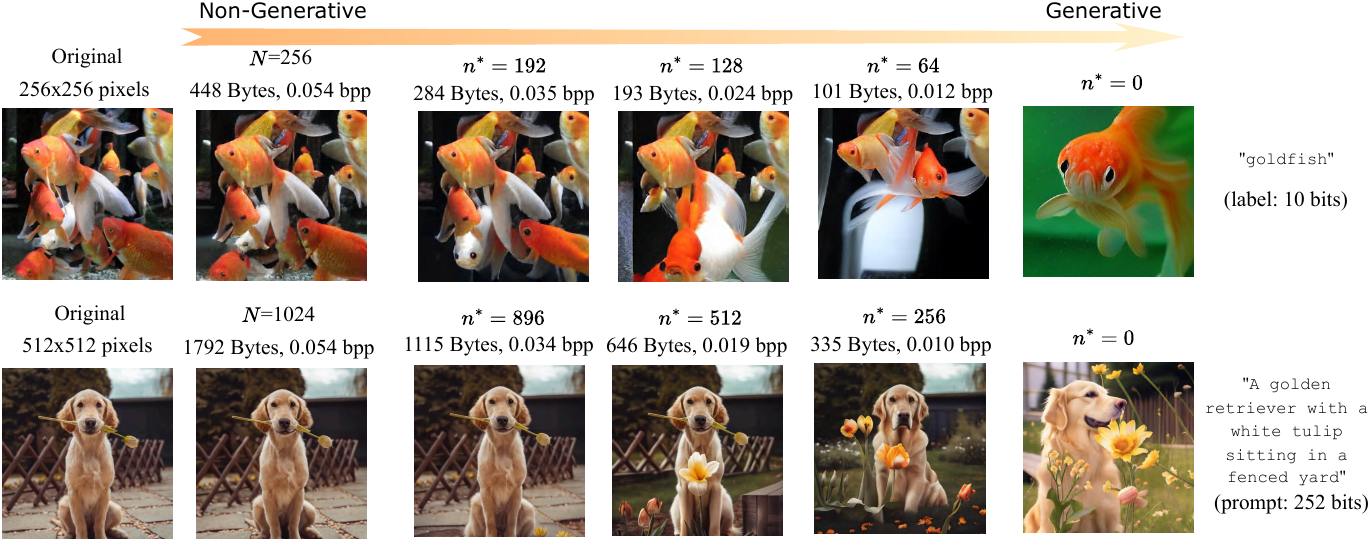}}

    \captionsetup{justification=justified, singlelinecheck=false}

    \caption{Visualization of image reconstruction at different token truncating levels for t2i and c2i scenarios.}
    \label{Fig.3}
    \vspace{-4mm}
\end{figure*}

As shown in Fig.~\ref{Fig.2}, Ada-TokenCom-\textit{cond} spans only $0.0062$--$0.0468$ bpp, which is over $90\%$ lower than conventional learning-based codecs such as Cheng2020-Attention ($\approx 0.145$--$0.711$ bpp), and also much lower than HiFiC ($0.688$--$1.003$ bpp). Despite this substantial rate reduction, Ada-TokenCom-\textit{cond} maintains high CLIP similarity scores of $0.923$--$0.973$. It also achieves favorable FID in the low-rate region, indicating strong semantic alignment and distribution-level realism.
Compared with vanilla TokenCom, which uses a fixed rate of $\log_2 V/256 = 0.0547$ bpp, Ada-TokenCom further reduces the bitrate through AR-based arithmetic coding and supports flexible rate adaptation via token truncation and adaptive token allocation. Conventional codecs still achieve higher PSNR, as expected, since they are optimized for pixel-level fidelity, whereas the discrete tokenizer and generative reconstruction in Ada-TokenCom do not preserve exact pixel values. Nevertheless, Ada-TokenCom achieves a competitive LPIPS range of $0.140$--$0.595$ at much lower rates, suggesting that it trades pixel-level accuracy for semantic consistency and perceptual plausibility in the ultra-low-rate regime.

To further illustrate the reconstruction behavior, Fig.~\ref{Fig.3} presents representative visual examples under different truncation lengths for both the c2i and text-to-image (t2i) settings. 
In both cases, the reconstructed images exhibit progressively improved visual quality and semantic consistency as $n^*$ increases. For the c2i example, the semantic condition is the class label ``goldfish'', whose transmission overhead is negligible ($\log_2 1000 \approx 10$ bits using fixed-length label encoding over ImageNet classes). For the t2i example, we use the $32\times32$-grid VQGAN tokenizer released with LlamaGen and pre-trained on the LAION-COCO dataset, together with the corresponding text-conditional AR model. The test image is taken from ImageNet, and the text condition is given by the automatically generated caption ``A golden retriever with a white tulip sitting in a fenced yard''. The prompt is compressed using character-frequency-based arithmetic coding, resulting in an overhead of 252 bits (31.5 Bytes), which is negligible compared with the transmitted image-token payload (e.g., 1115 Bytes in Fig. 4). Accordingly, the experiments focus on the transmission cost of image tokens only. Although the tokenizer resolution, pre-training dataset, and conditioning modality differ from those in the c2i setting, similar observations hold: reducing $n^*$ lowers both the resulting bpp and the number of transmitted bytes, while the reconstructed image still preserves the main semantics specified by the condition.

\subsection{Wireless Transmission Performance}
To evaluate the proposed framework under wireless transmission, the Lyapunov-based joint prefix-length and MCS adaptation formulated in~(\ref{eq:cross_layer_problem}) is simulated with semantic quality values obtained from actual end-to-end reconstructions. In each time slot, the full pipeline: source truncation, AC encoding, channel coding and modulation, fading-channel transmission, AC decoding, and AR-based token completion, is executed. The CLIP similarity between the reconstructed and original images is used as the semantic quality $A$.

The c2i model with image resolution $256\times256$ and class-label conditioning is adopted, and inference is performed on the ImageNet-V2 dataset. The candidate prefix lengths are $n^* \in \{32, 64, \ldots, 256\}$, aligned with the boundaries of the flattened latent grid of the VQGAN tokenizer. The corresponding semantic quality values, measured as average CLIP similarity over the evaluation set, range from $A = 0.843$ at $n^* = 32$ to $A = 0.937$ at $n^* = 256$, as listed in Table~\ref{tab:retained_tokens_A}. Five candidate channel coding and modulation schemes are predefined, i.e., $|\mathcal{M}| = 10$, covering uncoded transmission, rate-$1/2$, $2/3$, and $1/5$ convolutional codes, and a rate-$1/3$ turbo code, each paired with both QPSK and 16QAM. The corresponding PER model parameters $a_m$ and $b_m$ in~\eqref{eq:per_final} are summarized in Table~\ref{tab:a_b}. 
Simulations are conducted using NVIDIA Sionna \cite{sionna}, an open-source link-level simulation toolbox built on TensorFlow. The channel is assumed to remain constant within each time slot, and the channel coefficient $h_t$ is assumed to be known at the receiver.

\begin{table}[!tb]
\setlength{\tabcolsep}{3pt}
\centering
\caption{\centering {Candidate Sets of Channel Coding and Modulations Schemes}}
\label{tab:a_b}
\small
\begin{tabular}{c|c|c|cc|cc}
\hline
 & &&\multicolumn{2}{c|}{QPSK} & \multicolumn{2}{c}{16QAM} \\
Code & Rate & Polynominal & $\alpha_m$ & $\beta_m$ & $\alpha_m$ & $\beta_m$ \\
\hline
Uncoded & $1$ & $-$ & 1.742 & -2.188 & 8.662 & -13.306 \\
Conv Code &$1/2$  & $[5,7]_8$ & 0.320 & 0.011 & 1.103 & -0.277  \\
Conv Code &$2/3$  & $[51,31,13]_8$ & 0.157 & 0.225 & 0.506 & 0.623  \\
Conv Code &$1/5$  & \makecell{[35, 33, 27, \\ 37, 23]$_8$} & 0.086 & 0.070 & 0.268 & 0.075 \\
Turbo Code &$1/3$ & $[1,5/7,5/7]_8$ & 0.031 & 0.560 & 0.087 & 1.630  \\
\hline
\end{tabular}
\end{table}
\begin{table}[!tb]
\setlength{\tabcolsep}{3pt}
\centering
\caption{\centering {Semantic quality $A$ (CLIP) Under Different Prefix Lengths $n^*$}}
\label{tab:retained_tokens_A}
\small
\begin{tabular}{c|cccccccc}
\hline
\ $n^*$ & 256 & 224 & 192 & 160 & 128 & 96 & 64 & 32 \\
\hline
$A$ & 0.937 & 0.931 & 0.922 & 0.913 & 0.898 & 0.880 & 0.861 & 0.843 \\
\hline
\end{tabular}
\end{table}

\subsubsection{End-to-End Comparison with Baseline Schemes}
The proposed framework is compared with SwinJSCC, a state-of-the-art DeepJSCC framework based on the Swin Transformer that supports joint SNR and rate adaptation~\cite{10589474}. For a fair comparison, SwinJSCC is trained and validated on ImageNet-1K with 1.2M and 50K images, respectively, while inference for all methods is performed on the ImageNet-V2 dataset containing 10K images. The average channel bandwidth ratio is defined as $\text{CBR} = N_{\text{avg}} / (3 \times H \times W)$, where $N_{\text{avg}}$ is the average number of transmitted channel symbols over all time slots. Simulation is conducted over $T=10^3$ time slots. The overhead caused by retransmissions and transmission failures is included when computing the average CBR, while the reconstruction quality metrics are evaluated over successfully reconstructed images.

\begin{figure*}[!t]
	\centering
    {\centering\includegraphics[width=1\linewidth]{ 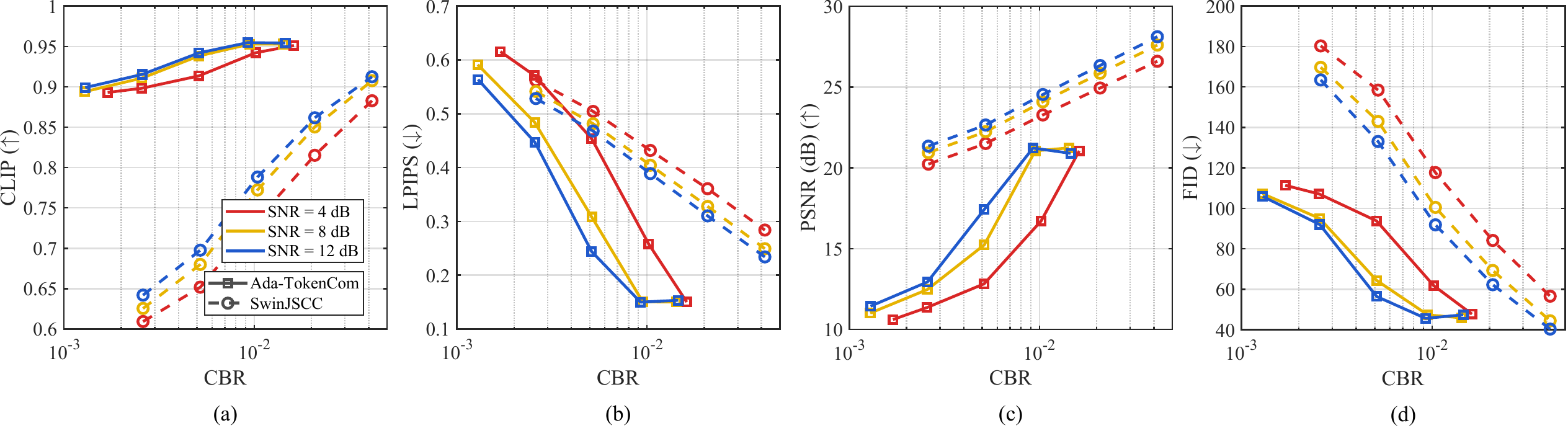}}  
    
    \captionsetup{justification=justified, singlelinecheck=false}
	\caption{Performance comparison of reconstruction among the proposed LlamaGen-based Ada-TokenCom framework, SwinJSCC with rate and SNR adaptation on the ImageNet-V2 dataset.} 
    
	\label{Fig.vs} 
    \vspace{-2mm}
\end{figure*}

\begin{figure}[!t]
    \centering
    \subfloat{\includegraphics[width=0.46\textwidth]{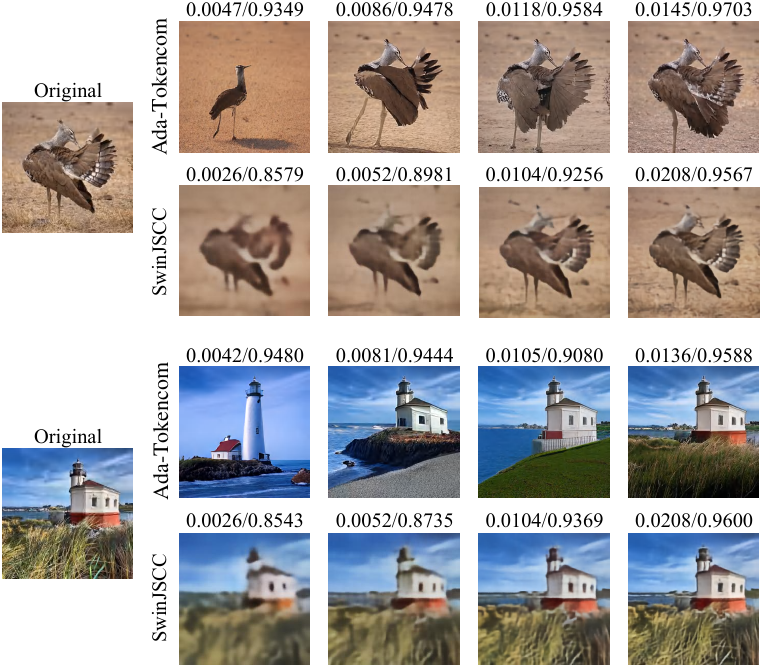}}
    \captionsetup{justification=justified, singlelinecheck=false}
    \caption{Visualization of image reconstruction at different channel bandwidth ratios for the Ada-TokenCom framework and SwinJSCC with rate and SNR adaptation at $\text{SNR}=12\text{dB}$. The value shown on each image denotes CBR / CLIP score.}
    \label{fig.vis_jscc}
     \vspace{-2mm}
\end{figure}
For the proposed framework, the operating CBR is adapted through the average channel-symbol budget $C_{\mathrm{th}}$ in the optimization problem~(\ref{eq:cross_layer_problem}), which is swept over discrete values ranging from 250 to 5000. A larger $C_{\mathrm{th}}$ permits more channel symbols per slot and thus a higher CBR, while a smaller $C_{\mathrm{th}}$ forces the optimizer to select shorter prefix lengths and more bandwidth-efficient MCS configurations.
The trade-off parameter is set to $\eta = 10^6$ which represents the empirically optimal point found via exhaustive grid search over $\{10^0,10^1,\ldots, 10^8\}$.
Recall from~(\ref{eq:per_slot_policy}) that $\eta$ governs the balance between semantic quality maximization and long-term resource expenditure: a small $\eta$ favors queue stability and conservative transmission, whereas a large $\eta$ aggressively pursues higher semantic quality at the cost of increased channel usage.

As illustrated in Fig.~\ref{Fig.vs}, we compare Ada-TokenCom with SwinJSCC under different channel conditions and bandwidth constraints. 
In Fig.~\ref{Fig.vs}(a), Ada-TokenCom achieves substantially higher CLIP similarity than SwinJSCC in the low-CBR regime, demonstrating stronger semantic preservation under limited wireless resources. Around CBR $\approx 10^{-2}$, Ada-TokenCom improves CLIP similarity by about $21\%$--$28\%$ across different SNRs, and the gain can reach approximately $47\%$ at lower CBR. In Fig.~\ref{Fig.vs}(b), Ada-TokenCom also yields lower LPIPS as the available bandwidth increases, indicating improved perceptual reconstruction quality. At comparable CBRs, it reduces LPIPS by up to about $60\%$.
Fig.~\ref{Fig.vs}(c) shows that SwinJSCC obtains higher PSNR, reflecting its advantage in pixel-level fidelity. Fig.~\ref{Fig.vs}(d) shows that Ada-TokenCom achieves markedly lower FID in the low-to-medium CBR range, with reductions of up to about $77\%$ at high SNR. These results indicate that Ada-TokenCom prioritizes semantic and perceptual fidelity under stringent bandwidth constraints, while SwinJSCC is more favorable when pixel-wise distortion, as measured by PSNR, is the primary objective.

Fig.~\ref{fig.vis_jscc} presents representative visual comparisons between Ada-TokenCom and SwinJSCC at $\text{SNR} = 12\,\text{dB}$. At comparable CBR, the proposed framework produces reconstructions that are semantically coherent with the original content, whereas SwinJSCC exhibits visible artifacts under low-bandwidth conditions. By leveraging pre-trained large models as implicit semantic codecs, the proposed framework decouples model training from channel adaptation, enabling reusable semantic communication across application domains without per-dataset end-to-end retraining.

\subsubsection{Validation of the Lyapunov-Based Adaptation Policy}
We further investigate how the proposed Lyapunov-based policy adapts online and clarify the role of the trade-off parameter $\eta$. Although Lyapunov drift-plus-penalty is a classical online optimization tool, its role here is to provide a lightweight semantic-rate adapter that jointly adapts the prefix length and MCS under a long-term symbol budget, without requiring non-causal knowledge of future SNRs.

We sweep $\eta$ under four dynamic SNR scenarios, each simulated independently over $T=10^4$ time slots. The proposed policy is compared with the following reference schemes:

\begin{itemize}
    \item \emph{Per-slot oracle:} a causal but myopic baseline that maximizes the instantaneous semantic quality under the current-slot symbol budget. It does not use future SNR information, does not maintain a virtual queue, and cannot borrow transmission budget across time.

    \item \emph{Long-term oracle:} an offline benchmark with non-causal knowledge of the entire SNR sequence. It optimizes the cumulative semantic quality under the same long-term average budget and therefore serves as an upper bound with strict budget constraint.

    \item \emph{Fixed:} two non-adaptive baselines that always use a fixed number of retained tokens, with $n=32$ and $n=256$, respectively, regardless of channel condition or queue state.
\end{itemize}

\vspace{-2mm}
\begin{figure}[t]
    \subfloat[]{\includegraphics[width=0.24\textwidth]{ 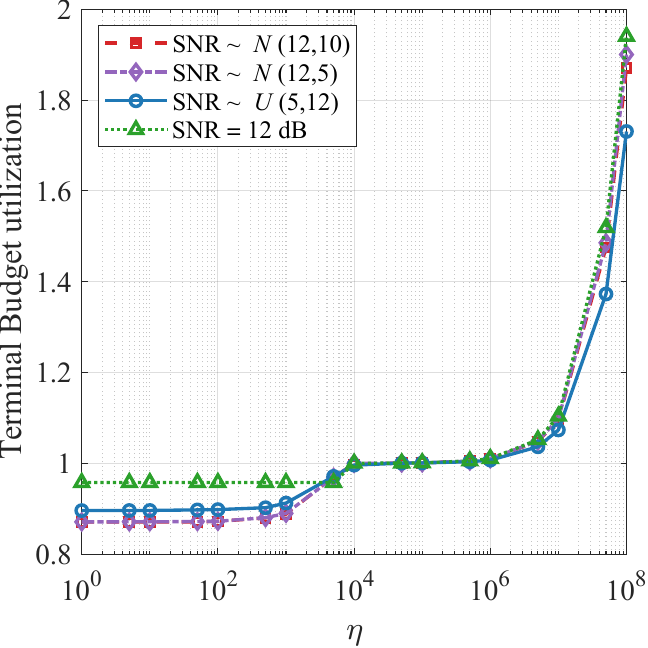}}
    \subfloat[]{\includegraphics[width=0.24\textwidth]{ 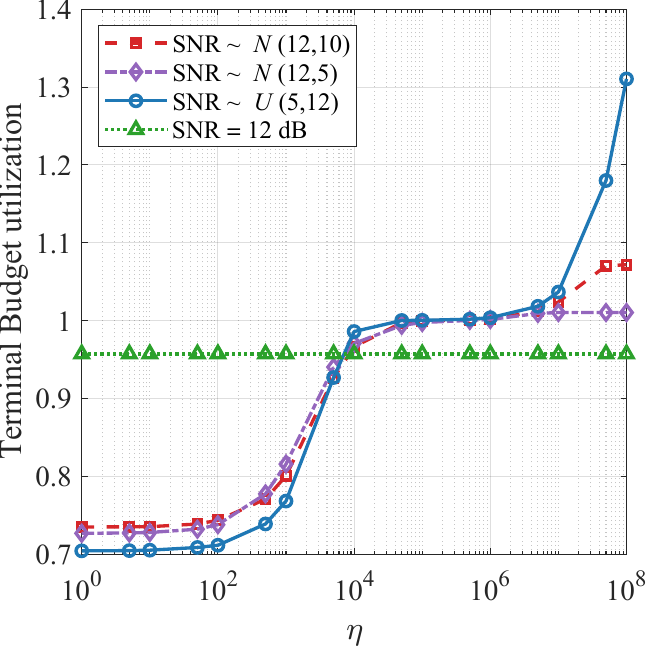}}

    \captionsetup{justification=justified, singlelinecheck=false}

    \caption{Effect of the Lyapunov weight~$\eta$ on terminal
  budget utilization at (a)~$C_{\mathrm{th}}=500$ and (b)~$C_{\mathrm{th}}=1500$ under four SNR
  regimes.}
    \label{fig:reward_queue}
\end{figure}

\begin{figure}[!t]
    \centering
    \subfloat{\includegraphics[width=0.5\textwidth]{ 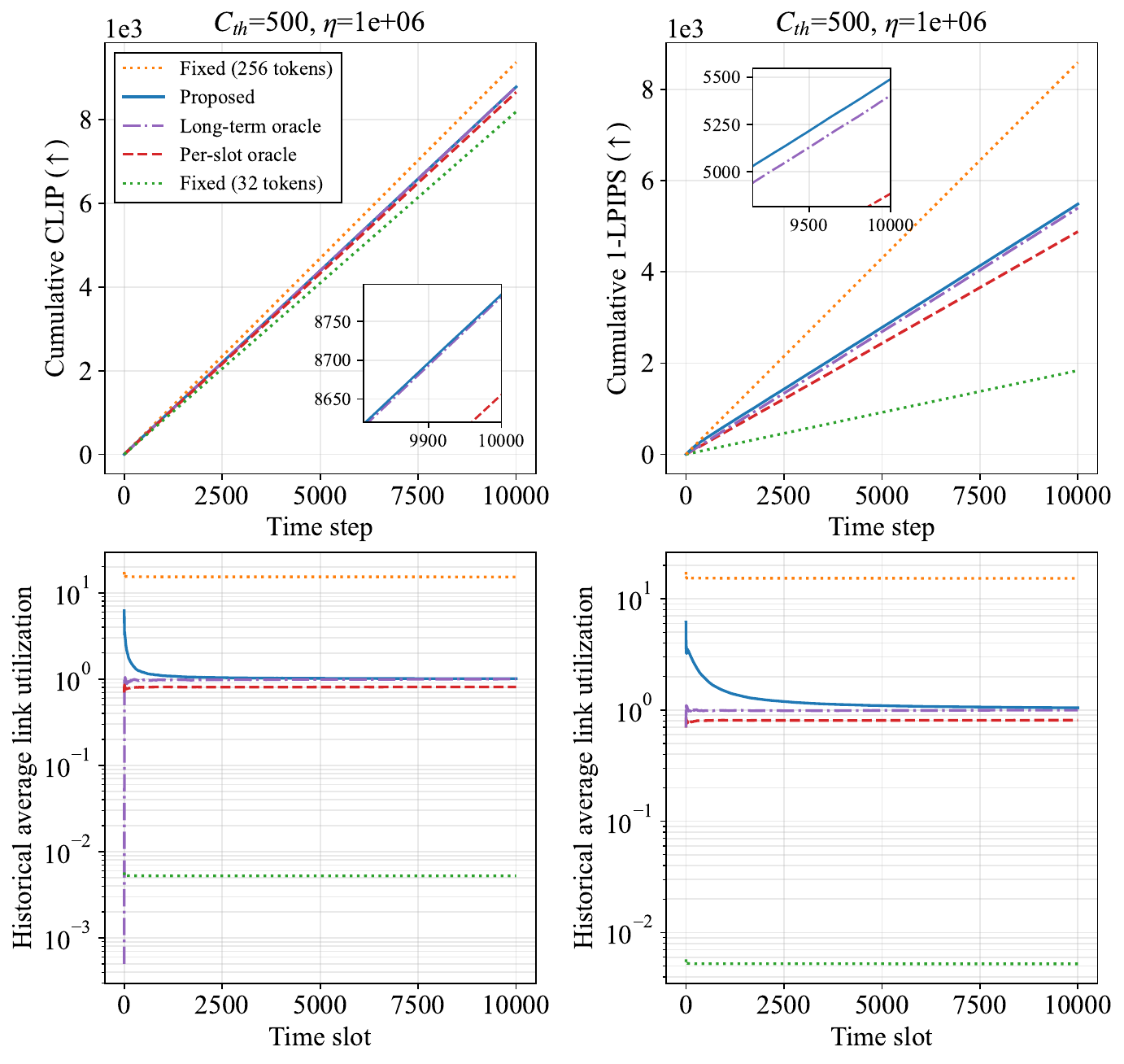}} 

    \captionsetup{justification=justified, singlelinecheck=false}

      \caption{Per-slot cumulative CLIP and $(1-\mathrm{LPIPS})$ (upper row) with their
  historical average link utilization (lower row) under
  $\mathrm{SNR}\sim N(12,10)$ when $C_\mathrm{th}=500$.}
    \label{fig:traces_grid}
\end{figure}

Fig.~\ref{fig:reward_queue} shows the terminal budget utilization under different values of $\eta$. A larger $\eta$ gives more weight to semantic quality in the drift-plus-penalty objective and therefore makes the adapter more aggressive in selecting high-quality and high-symbol-consumption actions. Consequently, small $\eta$ values lead to under-utilization of the available budget, whereas excessively large values may cause visible budget violations. Across the considered SNR regimes and budget levels, $\eta\approx10^6$ provides a representative aggressive yet near-feasible operating point, and is therefore used in the following end-to-end comparisons.

\begin{table*}[!t]
\centering
\caption{Comparison of model size, inference cost, and computation latency\textsuperscript{1}.}
\label{tab:complexity_latency}
\small

\setlength{\tabcolsep}{4pt}

\begin{tabularx}{\textwidth}{l c >{\centering\arraybackslash}X c c}
\toprule
\textbf{Method} &
\textbf{\makecell[c]{Parameters / M \\ (Encoder + Decoder)}} &
\textbf{\makecell[c]{Inference Cost / GFlops\\ (Encoder + Decoder)}} &
\textbf{Encoder Latency} &
\textbf{Decoder Latency} \\
\midrule

SwinJSCC &
18.34 + 14.73 &
35.01 + 34.11 &
10 ms &
6 ms \\

HiFiC &
24.69 + 168.28 &
15.28 + 116.47 &
83 ms &
188 ms \\

BPG &
-- &
-- &
307--824 ms &
29--43 ms \\

\midrule

Ada-TokenCom (LlamaGen-L)\textsuperscript{2} &
372.21 + 385.36  &
311.41 + 422.47 &
19 ms &
1012 ms \\

\bottomrule
\end{tabularx}

\vspace{2pt}
\begin{minipage}{0.95\textwidth}
\footnotesize
\textsuperscript{1} Latencies are measured using an NVIDIA 5090 GPU and an Intel Xeon Platinum 8260 CPU. BPG computations are primarily CPU-based.\\
\textsuperscript{2} For Ada-TokenCom, the encoder consists of a 29.30 M-parameter tokenizer and a 342.91 M-parameter AR model, while the decoder consists of a 42.45 M-parameter tokenizer and a 342.91 M-parameter AR model. The corresponding inference costs are 138.54 G + 172.87 G and 169.66 G + 252.81 G FLOPs, respectively.
\end{minipage}
\vspace{-5mm}
\end{table*}

Fig.~\ref{fig:traces_grid} further illustrates the temporal behavior of the proposed policy under a representative SNR process with $\gamma_t\sim N(12,10)$. $\eta$ is selected as the largest value that satisfies the terminal long-term budget constraint. The proposed policy achieves a final-slot throughput close to that of the long-term oracle, while requiring no non-causal information.
Under limited budgets, the proposed policy attains higher cumulative CLIP and cumulative perceptual reward, where the latter is measured by $(1-\mathrm{LPIPS})$ so that larger values correspond to better perceptual quality. Note that no negative $(1-\mathrm{LPIPS})$ values were found during testing.
The link utilization of the proposed policy may temporarily exceed one in the early slots, but it gradually converges to the constraint boundary as the virtual queue accumulates penalty. In contrast, the fixed policy with $n=32$ is feasible but overly conservative, while the fixed policy with $n=256$ achieves high quality at the cost of persistent budget violation. These two fixed baselines represent the lower and upper performance bounds achievable by layer-separation-based schemes without online prefix-length and MCS adaptation.

\vspace{-3mm}
\subsection{Computational Complexity and Latency Analysis}

As shown in~(\ref{eq:recon}), the proposed framework reduces the number of transmitted bits through truncation, but it does not reduce the total number of receiver-side AR inference steps. For a token sequence of length $N$ with retained prefix length $n^*$, the shared AR model is invoked during the first $n^*$ steps to recover the arithmetic-coded prefix and during the remaining $(N-n^*)$ steps to generate the missing tail tokens. Hence, the decoder still performs $N$ sequential model evaluations in total, regardless of the truncation length. In this sense, the proposed method has the same order of decoding complexity as full token transmission with arithmetic coding, while enabling semantic completion of the discarded tail tokens. 

At the transmitter, all retained prefix tokens are available. Their conditional probabilities can therefore be computed in a single prefill process, rather than through $n^*$ token-by-token AR iterations. This removes the sequential prediction loop for the retained prefix and substantially reduces the encoding latency. In our implementation, the LlamaGen forward pass is additionally accelerated with PyTorch \texttt{torch.compile}, which fuses and optimizes the model execution graph during inference. This compilation is used as a runtime optimization and does not change the probability distributions or the transmitted bitstream.

Table~\ref{tab:complexity_latency} summarizes the model size, inference cost, and measured latency of the proposed method and the baseline models. Autoencoder-based baselines such as SwinJSCC and HiFiC are computationally lighter, whereas the proposed AR framework incurs a higher decoding latency due to full-horizon sequential inference. Nevertheless, this additional computation is exchanged for substantially improved compression efficiency and semantic reconstruction quality in the low-rate regime. 
For Ada-TokenCom, 
we report the total latency as
$
T_{\mathrm{tokenizer}}+T_{\mathrm{AR}} + T_{\mathrm{AC}},
$
where $T_{\mathrm{tokenizer}}$ comes from tokenization or detokenization process, $T_{\mathrm{AR}}$ comes from AR probability estimation, and $T_{\mathrm{AC}}$ includes both softmax-based probability computation and AC coding. 
The encoder latency is $T_{\mathrm{tokenizer}}=7.1$ ms, $T_{\mathrm{AR}}=3.6$ ms, and $T_{\mathrm{AC}}=8.3$ ms, where $T_{\mathrm{AC}}$ consists of 4.1 ms for softmax computation, and 4.2 ms for AC coding. The full-tokens prefill enables efficient encoder-side computation.
At the decoder, the measured latency is $T_{\mathrm{tokenizer}}=8.4$ ms, $T_{\mathrm{AR}}=908.1$ ms, and $T_{\mathrm{AC}}=95.3$ ms, including 82.0 ms for softmax computation and 13.3 ms for AC coding. The higher AR latency stems from enforcing exact numerical consistency with arithmetic coding. Specifically, to ensure identical probability evaluations between encoder and decoder in simulation, the full-prefix AR distribution is recomputed at each decoding step instead of standard token-by-token inference. Consequently, decoding a 256-token sequence requires 256 repeated prefill evaluations, which dominates the overall latency. This design is intentionally conservative to guarantee strict encoder-–decoder consistency. In principle, full-sequence prefill and incremental AR decoding are equivalent, leaving substantial room for optimization.

\vspace{-4mm}

\section{Conclusion and Future Work}

This paper proposed \textit{Ada-TokenCom}, a rate-adaptive digital TokenCom framework that jointly integrates autoregressive token compression, receiver-side token generation, and Lyapunov-based cross-layer adaptation. Extensive experiments demonstrated that the proposed framework achieved an effective trade-off between semantic reconstruction quality and communication efficiency under bandwidth-constrained wireless environments.

Ada-TokenCom is particularly suitable for bandwidth-limited applications in which receiver-side computation can be traded for reduced communication overhead. A current limitation is the additional latency introduced by AR token prediction. Future work will therefore investigate faster generation mechanisms, the joint communication and generation tradeoff, and coarse-to-fine token representations that progressively refine reconstruction quality. The framework can also be extended to other modalities with structured token dependencies, such as video and multimodal data, which we leave as future work.

\setcounter{equation}{0}
\renewcommand{\theequation}{a.\arabic{equation}}

\balance
\bibliographystyle{IEEEtran}
\bibliography{ref}

@inproceedings{qiao2025todma,
  author    = {Li Qiao and Mahdi Boloursaz Mashhadi and Zhen Gao and Deniz G{\"u}nd{\"u}z},
  title     = {Token-Domain Multiple Access: Exploiting Semantic Orthogonality for Collision Mitigation},
  booktitle = {Proc. IEEE Conf. Comput. Commun. Workshops (INFOCOM WKSHPS)},
  pages     = {1--6},
  year      = {2025},
  doi       = {10.1109/INFOCOMWKSHPS65812.2025.11152964}
}

@article{qiao2025tokencom,
  author  = {Li Qiao and Mahdi Boloursaz Mashhadi and Zhen Gao and Rahim Tafazolli and Mehdi Bennis and Dusit Niyato},
  title   = {Token Communications: A Large Model-Driven Framework for Cross-Modal Context-Aware Semantic Communications},
  journal = {IEEE Wireless Commun.},
  volume  = {32},
  number  = {5},
  pages   = {80--88},
  year    = {2025},
  doi     = {10.1109/MWC.001.2500084}
}

@misc{men2026video,
  author = {Jingxuan Men and others},
  title  = {Video {TokenCom}: Textual Intent-Guided Multi-Rate Video Token Communications with {UEP}-Based Adaptive Source-Channel Coding},
  howpublished = {arXiv preprint arXiv:2603.02470},
  year   = {2026}
}

@article{wang2025resicomp,
  author  = {Sixian Wang and Jincheng Dai and Xiaoqi Qin and Ke Yang and Kai Niu and Ping Zhang},
  title   = {{ResiComp}: Loss-Resilient Image Compression via Dual-Functional Masked Visual Token Modeling},
  journal = {IEEE Trans. Circuits Syst. Video Technol.},
  volume  = {35},
  number  = {7},
  pages   = {7181--7195},
  year    = {2025},
  doi     = {10.1109/TCSVT.2025.3539747}
}

@article{wei2026token,
  author  = {Hao Wei and Wanli Ni and Wen Wang and Wenjun Xu and Dusit Niyato and Ping Zhang},
  title   = {Token Communication in the Era of Large Models: An Information Bottleneck-Based Approach},
  journal = {IEEE Wireless Commun. Lett.},
  volume  = {15},
  pages   = {186--190},
  year    = {2026},
  doi     = {10.1109/LWC.2025.3622628}
}

@misc{lee2025low,
  author       = {Seunghun Lee and Jihong Park and Jinho Choi and Hyuncheol Park},
  title        = {Low-Complexity Semantic Packet Aggregation for Token Communication via Lookahead Search},
  howpublished = {arXiv preprint arXiv:2506.19451},
  year         = {2025}
}

@article{zhang2026task,
  author  = {Junhe Zhang and Wanli Ni and Pengwei Wang and Dongyu Wang},
  title   = {Task-Oriented Multimodal Token Transmission in Resource-Constrained Multiuser Networks},
  journal = {IEEE Wireless Commun. Lett.},
  volume  = {15},
  pages   = {570--574},
  year    = {2026},
  doi     = {10.1109/LWC.2025.3628928}
}

@misc{zeinali2026wireless,
  author       = {Farshad Zeinali and Mahdi Boloursaz Mashhadi and Dusit Niyato and Rahim Tafazolli},
  title        = {Wireless {TokenCom}: {RL}-Based Tokenizer Agreement for Multi-User Wireless Token Communications},
  howpublished = {arXiv preprint arXiv:2602.12338},
  year         = {2026}
}

@misc{jiang2026tokencom,
  author       = {Feibo Jiang and Siwei Tu and Li Dong and Xiaolong Li and Kezhi Wang and Cunhua Pan and Zhu Han and Jiangzhou Wang},
  title        = {{TokenCom}: Vision-Language Model for Multimodal and Multitask Token Communications},
  howpublished = {arXiv preprint arXiv:2603.00482},
  year         = {2026}
}

@article{zhang2026tokencomuep,
  author  = {Kaizheng Zhang and Zuolin Jin and Yi Zhang and Zhihang Cheng and Ming Zeng and Li Qiao and Zesong Fei},
  title   = {{TokenCom-UEP}: Semantic Importance-Matched Unequal Error Protection for Resilient Image Transmission},
  journal = {IEEE Wireless Commun. Lett.},
  volume  = {15},
  pages   = {2809--2813},
  year    = {2026},
  doi     = {10.1109/LWC.2026.3683938}
}

@article{ying2026joint,
  author  = {Jingkai Ying and Zhijin Qin and Yulong Feng and Liejun Wang and Xiaoming Tao},
  title   = {Joint Semantic-Channel Coding and Modulation for Token Communications},
  journal = {IEEE Trans. Wireless Commun.},
  volume  = {25},
  pages   = {8179--8193},
  year    = {2026},
  doi     = {10.1109/TWC.2025.3636174}
}

@article{devoto2026adaptive,
  author  = {Alessio Devoto and others},
  title   = {Adaptive Semantic Token Communication for Transformer-Based Edge Inference},
  journal = {IEEE Trans. Mach. Learn. Commun. Netw.},
  volume  = {4},
  pages   = {422--437},
  year    = {2026},
  doi     = {10.1109/TMLCN.2026.3659819}
}

@article{erak2026adaptive,
  author  = {Omar Erak and Omar Alhussein and Hatem Abou-Zeid and Mehdi Bennis and Sami Muhaidat},
  title   = {Adaptive Token Merging for Efficient Transformer Semantic Communication at the Edge},
  journal = {IEEE Open J. Commun. Soc.},
  volume  = {7},
  pages   = {4112--4128},
  year    = {2026},
  doi     = {10.1109/OJCOMS.2026.3676928}
}

@article{yin2024survey,
  author  = {Shukang Yin and Chaoyou Fu and Sirui Zhao and Ke Li and Xing Sun and Tong Xu and Enhong Chen},
  title   = {A Survey on Multimodal Large Language Models},
  journal = {Natl. Sci. Rev.},
  volume  = {11},
  number  = {12},
  pages   = {nwae403},
  year    = {2024}
}

@article{chen2025communication,
  author  = {Xuyang Chen and others},
  title   = {The Communication and Computation Trade-Off in Wireless Semantic Communications},
  journal = {IEEE Wireless Commun. Lett.},
  volume  = {14},
  number  = {7},
  pages   = {2259--2263},
  year    = {2025},
  doi     = {10.1109/LWC.2025.3569205}
}

@article{dai2022nonlinear,
  author  = {Jincheng Dai and others},
  title   = {Nonlinear Transform Source-Channel Coding for Semantic Communications},
  journal = {IEEE J. Sel. Areas Commun.},
  volume  = {40},
  number  = {8},
  pages   = {2300--2316},
  year    = {2022}
}

@article{huang2025visual,
  title={Visual fidelity index for generative semantic communications with critical information embedding},
  author={Huang, Jianhao and Zeng, Qunsong and Huang, Kaibin},
  journal={arXiv preprint arXiv:2505.10405},
  year={2025}
}

@ARTICLE{11112664,
  author={Yin, Hang and others},
  journal={IEEE Trans. Veh. Technol.}, 
  title={Generative Video Semantic Communication via Multimodal Semantic Fusion With Large Model}, 
  year={2026},
  volume={75},
  number={1},
  pages={1701-1706},
  doi={10.1109/TVT.2025.3595688}}

@ARTICLE{10599525,
  author={Qiao, Li and others},
  journal={IEEE Wireless Commun. Lett.}, 
  title={Latency-Aware Generative Semantic Communications With Pre-Trained Diffusion Models}, 
  year={2024},
  volume={13},
  number={10},
  pages={2652-2656},
  doi={10.1109/LWC.2024.3429295}}

@article{bourtsoulatze2019deep,
  title={Deep joint source-channel coding for wireless image transmission},
  author={Bourtsoulatze, Eirina and others},
  journal={IEEE Trans. Cogn. Commun. Netw.},
  volume={5},
  number={3},
  pages={567--579},
  year={2019},
  publisher={IEEE}
}

@article{ren2025separate,
  author  = {Tianqi Ren and others},
  title   = {Separate Source Channel Coding Is Still What You Need: An {LLM}-Based Rethinking},
  journal = {ZTE Commun.},
  volume  = {23},
  number  = {1},
  pages   = {30--44},
  month   = mar,
  year    = {2025}
}

@inproceedings{deletang2024language,
  author    = {Gregoire Deletang and others},
  title     = {Language Modeling Is Compression},
  booktitle = {Int. Conf. Learn. Represent.},
  year      = {2024}
}

@article{sun2024autoregressive,
  title={Autoregressive Model Beats Diffusion: Llama for Scalable Image Generation},
  author={Sun, Peize and others},
  journal={arXiv preprint arXiv:2406.06525},
  year={2024}
}

@InProceedings{pmlr-v139-radford21a,
  author    = {Alec Radford and others},
  title     = {Learning Transferable Visual Models From Natural Language Supervision},
  booktitle = {Int. Conf. Mach. Learn.},
  volume    = {139},
  pages     = {8748--8763},
  year      = {2021}
}

@article{balle2018variational,
  title={Variational image compression with a scale hyperprior},
  author={Ball{\'e}, Johannes and others},
  journal={arXiv preprint arXiv:1802.01436},
  year={2018}
}

@article{begaint2020compressai,
  title={{CompressAI}: {A} PyTorch library and evaluation platform for end-to-end compression research},
  author={B{\'e}gaint, Jean and others},
  year={2020},
  journal={arXiv preprint arXiv:2011.03029}
}

@article{mentzer2020high,
  title={High-Fidelity Generative Image Compression},
  author={Mentzer, Fabian and others},
  journal={Adv. Neural Inf. Process. Syst.},
  volume={33},
  year={2020}
}

@inproceedings{zhang2018unreasonable,
  author    = {Richard Zhang and others},
  title     = {The Unreasonable Effectiveness of Deep Features as a Perceptual Metric},
  booktitle = {IEEE Conf. Comput. Vis. Pattern Recognit. (CVPR)},
  pages     = {586--595},
  year      = {2018}
}

@inproceedings{NIPS2017_8a1d6947,
 author = {Heusel, Martin and others},
 booktitle = {Adv. Neural Inf. Process. Syst.},
 title = {{GANs} Trained by a Two Time-Scale Update Rule Converge to a Local Nash Equilibrium},
 volume = {30},
 year = {2017}
}

@ARTICLE{6514951,
  author={Wu, Jingxian and others},
  journal={IEEE J. Sel. Areas Commun.}, 
  title={Energy Efficiency and Spectral Efficiency Tradeoff in {Type-I} {ARQ} Systems}, 
  year={2014},
  volume={32},
  number={2},
  pages={356-366},
  doi={10.1109/JSAC.2014.141215}
}

@ARTICLE{10982132,
  author={Ren, Mengmeng and Qiao, Li and others},
  journal={IEEE Trans. Veh. Technol.}, 
  title={Generative Semantic Communication via Textual Prompts: Latency Performance Tradeoffs}, 
  year={2025},
  volume={74},
  number={9},
  pages={14843-14848},
  doi={10.1109/TVT.2025.3566488}}

@ARTICLE{10589474,
  author={Yang, Ke and others},
  journal={IEEE Trans. Cogn. Commun. Netw.}, 
  title={{SwinJSCC}: Taming Swin Transformer for Deep Joint Source-Channel Coding}, 
  year={2025},
  volume={11},
  number={1},
  pages={90-104},
  doi={10.1109/TCCN.2024.3424842}}

@software{sionna,
 title = {Sionna},
 author = {Hoydis, Jakob and others},
 note = {https://nvlabs.github.io/sionna/},
 year = {2022},
 version = {1.2.1}
}

@inproceedings{recht2019imagenet,
  author    = {Benjamin Recht and others},
  title     = {Do {ImageNet} Classifiers Generalize to {ImageNet}?},
  booktitle = {Int. Conf. Mach. Learn.},
  pages     = {5389--5400},
  year      = {2019}
}

@INPROCEEDINGS{4558591,
  author    = {Ioannis Chatzigeorgiou and others},
  title     = {On the Frame Error Rate of Transmission Schemes on Quasi-Static Fading Channels},
  booktitle = {Annu. Conf. Inf. Sci. Syst. (CISS)},
  pages     = {577--581},
  year      = {2008},
  doi       = {10.1109/CISS.2008.4558591}
}

@article{ToDMA,
  title={{ToDMA}: Large Model-Driven Token-Domain Multiple Access for Semantic Communications},
  author={Li Qiao and Mahdi Boloursaz Mashhadi and Zhen Gao and Robert Schober and Deniz Gündüz},
  journal={arXiv preprint arXiv:2505.10946v2},
  year={2025}
}

@ARTICLE{10960413,
  author={Xu, Chunmei and others},
  journal={IEEE J. Sel. Areas Commun.}, 
  title={Generative Semantic Communications With Foundation Models: Perception-Error Analysis and Semantic-Aware Power Allocation}, 
  year={2025},
  volume={43},
  number={7},
  pages={2493-2505},
  doi={10.1109/JSAC.2025.3559120}}

@inproceedings{Esser2021CVPR,
  author    = {P. Esser and others}, 
  title     = {Taming Transformers for High-Resolution Image Synthesis},
  booktitle = {Proc. IEEE/CVF Conf. Comput. Vis. Pattern Recognit.}, 
  month     = {Jun.},
  year      = {2021},
  pages     = {12873--12883}
}

@inproceedings{
miwa2026onedpiece,
title={{One-D-Piece}: Image Tokenizer Meets Quality-Controllable Compression},
author={Keita Miwa and others},
booktitle={Tokenization Workshop},
year={2026},
}

@inproceedings{
tian2024visual,
title={Visual Autoregressive Modeling: Scalable Image Generation via Next-Scale Prediction},
author={Keyu Tian and others},
booktitle={Adv. Neural Inf. Process. Syst.},
year={2024}
}

@ARTICLE{SemCom1,
  author={Gündüz, Deniz and others},
  journal={IEEE J. Sel. Areas Commun.}, 
  title={Beyond Transmitting Bits: Context, Semantics, and Task-Oriented Communications}, 
  year={2023},
  volume={41},
  number={1},
  pages={5-41},
  doi={10.1109/JSAC.2022.3223408}}

@ARTICLE{SemCom2,
  author={Yang, Wanting and others},
  journal={IEEE Commun. Surveys Tuts.}, 
  title={Semantic Communications for Future Internet: Fundamentals, Applications, and Challenges}, 
  year={2023},
  volume={25},
  number={1},
  pages={213-250},
  doi={10.1109/COMST.2022.3223224}}

@article{wangMultimodalLearningNexttoken2026,
  title = {Multimodal Learning with Next-Token Prediction for Large Multimodal Models},
  author = {Wang, Xinlong and others},
  year = 2026,
  month = jan,
  journal = {Nature},
  issn = {1476-4687},
  doi = {10.1038/s41586-025-10041-x},
}

@ARTICLE{10960324,
  author={Wang, Sixian and others},
  journal={IEEE J. Sel. Areas Commun.}, 
  title={{DiffCom}: Channel Received Signal Is a Natural Condition to Guide Diffusion Posterior Sampling}, 
  year={2025},
  volume={43},
  number={7},
  pages={2651-2666},
  doi={10.1109/JSAC.2025.3559158}}

@ARTICLE{11370276,
  author={Liu, Xinkai and others},
  journal={IEEE Trans. Veh. Technol.}, 
  title={Communicate Less, Synthesize the Rest: Latency-aware Intent-based Generative Semantic Multicasting with Diffusion Models}, 
  year={2026},
  volume={},
  number={},
  pages={1-16},
  doi={10.1109/TVT.2026.3660013}}

@InProceedings{Cheng_2020_CVPR,
author = {Cheng, Zhengxue and Sun, Heming and Takeuchi, Masaru and Katto, Jiro},
title = {Learned Image Compression With Discretized Gaussian Mixture Likelihoods and Attention Modules},
booktitle = {Proc. IEEE/CVF Conf. Comput. Vis. Pattern Recognit.},
month = {June},
year = {2020}
}

@article{wang2023wireless,
  author  = {Sixian Wang and others},
  title   = {Wireless Deep Video Semantic Transmission},
  journal = {IEEE J. Sel. Areas Commun.},
  volume  = {41},
  number  = {1},
  pages   = {214--229},
  year    = {2023},
  doi     = {10.1109/JSAC.2022.3221977}
}

@article{zhang2026semantics,
  author  = {Maojun Zhang and others},
  title   = {Semantics-Guided Diffusion for Deep Joint Source-Channel Coding in Wireless Image Transmission},
  journal = {IEEE Trans. Wireless Commun.},
  volume  = {25},
  pages   = {1547--1564},
  year    = {2026},
  doi     = {10.1109/TWC.2025.3591456}
}

@article{huang2025d2jscc,
  author  = {Jianhao Huang and others},
  title   = {{D$^2$-JSCC}: Digital Deep Joint Source-Channel Coding for Semantic Communications},
  journal = {IEEE J. Sel. Areas Commun.},
  volume  = {43},
  number  = {4},
  pages   = {1246--1261},
  year    = {2025},
  doi     = {10.1109/JSAC.2025.3531546}
}

@book{Neely2010,
  author    = {Michael J. Neely},
  title     = {Stochastic Network Optimization with Application to Communication and Queueing Systems},
  publisher = {Morgan \& Claypool Publishers},
  year      = {2010}
}

@ARTICLE{linarq,
  author={Shu Lin and others},
  journal={IEEE Commun. Mag.}, 
  title={Automatic-repeat-request error-control schemes}, 
  year={1984},
  volume={22},
  number={12},
  pages={5-17},
  doi={10.1109/MCOM.1984.1091865}}

@article{Witten1987,
  author  = {Ian H. Witten and others},
  title   = {Arithmetic Coding for Data Compression},
  journal = {Commun. ACM},
  volume  = {30},
  number  = {6},
  pages   = {520--540},
  year    = {1987}
}

\end{document}